\documentclass[11pt]{article}

\newcommand{\commentstarts}{\begin{centering}
\hspace{-1pt}\vrule\vrule
\begin{minipage}[t]{0.03\linewidth}
\hspace{0.025\linewidth}
\end{minipage}
\begin{minipage}[t]{0.95\linewidth}}
\newcommand{\commentends}{\end{minipage}
\end{centering}
\vspace{7pt}
}

\usepackage{graphicx}
\usepackage{alltt}
\usepackage{amsmath}
\usepackage{amssymb}
\usepackage{hyperref}

\newcounter{Theorems}
\newcounter{Definitions}
\newcounter{Conjectures}
\begin{document} 

\begin{titlepage}
\begin{flushright}

\end{flushright}

\begin{center}
{\Large\bf $ $ \\ $ $ \\
Deformations, renormgroup, symmetries, AdS/CFT
}\\
\bigskip\bigskip\bigskip
{\large Andrei Mikhailov${}^{\dag}$}
\\
\bigskip\bigskip
{\it Instituto de F\'{i}sica Te\'orica, Universidade Estadual Paulista\\
R. Dr. Bento Teobaldo Ferraz 271, 
Bloco II -- Barra Funda\\
CEP:01140-070 -- S\~{a}o Paulo, Brasil\\
}

\vskip 1cm
\end{center}

\begin{abstract}
   We consider the deformations of a supersymmetric quantum field theory by
   adding spacetime-dependent terms to the action. We propose to describe the
   renormalization of such deformations in terms of some cohomological
   invariants, a class of solutions of a Maurer-Cartan equation. We consider
   the strongly coupled limit of $N=4$ supersymmetric Yang-Mills theory.
   In the context of AdS/CFT correspondence, we explain what corresponds
   to our invariants in classical supergravity. There is a leg amputation procedure,
   which constructs a solution of the Maurer-Cartan equation from tree diagramms of SUGRA.
   We consider a particular example of the beta-deformation.  It is
   known that the leading term of the beta-function is cubic in the parameter
   of the beta-deformation. We give a cohomological interpretation of this
   leading term. We conjecture that it is actually encoded in some simpler
   cohomology class, which is quadratic in the parameter of the
   beta-deformation.
\end{abstract}

\vfill
{\renewcommand{\arraystretch}{0.8}%
\begin{tabular}{rl}
${}^\dag\!\!\!\!$ 
& 
\footnotesize{on leave from Institute for Theoretical and 
Experimental Physics,}
\\    
&
\footnotesize{NRC ``Kurchatov Institute'', Moscow, Russia}
\\
\end{tabular}
}

\end{titlepage}

\tableofcontents

\section{Introduction}

\subsection{Renormalization of the deformations of the action}\label{sec:IntroRenorm}

Consider a quantum field theory with an action $S$ invariant under some Lie algebra of symmetries $\bf g$.
Let us study its infinitesimal deformations of the theory, corresponding to the deformations of the action:
\begin{equation}\label{Deformations}
\delta S \;=\; \epsilon\int d^dx \sum_I f_I(x) U_I(x)
\end{equation}
where $\epsilon$ is an infinitesimal parameter, $f_I(x)$ are some space-time-dependent coupling constants
and $\{U_I\}$ is some set of local operators,
closed under $\bf g$ in the sense that the expressions on the RHS of Eq. (\ref{Deformations})
form a linear representations of $\bf g$. We call $T_0$ the linear space of this representation
\begin{equation}
T_0 \; = \;  \mbox{\tt\small linear space generated by } \int d^dx f_I(x) U_I(x)
\end{equation}
In principle, we can take  $\{U_I\}$ the set of all local operators of the theory.
But there could be smaller $\bf g$-invariant subspaces.

We can study the effects on the correlation functions, or perhaps on the $S$-matrix,
of the deformation of the form (\ref{Deformations}), to the linear order in $\epsilon$.
We can also study the effects
of the deformation (\ref{Deformations}) beyond linear order in $\epsilon$,
but this requires taking care of the definitions. To define (\ref{Deformations}), we expand in powers of $\epsilon$,
bringing down from the exponential expressions like:
\begin{equation}\label{MultipleIntegral}
\epsilon^n \int d^dx fU\cdots\int d^dx fU
\end{equation}
This has to be regularized, because of singularities due to collisions of $U$.
Suppose that the set $\{U_I\}$ is big enough in the sense that all the required
counterterms are linear combinations of $\{U_I\}$. The counterterms are not unique,
because we can always add a finite expression. Suppose that we have choosen some rule to fix this ambiguity.
Then, we have a map, parameterized by a small parameter $\epsilon$:
\begin{equation}\label{MapFinFieldTheoryContext}
F_{\epsilon}\;:\;T_0 \longrightarrow \left[\mbox{\tt\small space of finite deformations}\right]
\end{equation}

\subsection{Symmetries of undeformed theory act on deformations}\label{sec:ActionOfGOnDefs}

The space of finite deformations is not, in any useful sense, a linear space.
It is a ``nonlinear infinite-dimensional manifold''.
But it naturally comes with an {\em action} of $\bf g$.
Indeed, the  regularized expression
(\ref{MultipleIntegral}) is, in particular, a (non-local) operator in the original theory.
As the symmetry group of the undeformed theory acts on operators, it therefore
acts on deformations, bringing one deformation to another.

\vspace{10pt}
\noindent
Because we had some freedom in the choice of regularization, the map $F_{\epsilon}$ does
not necessarily commute with the action of $\bf g$.
Can we choose regularization
with some care, so the resulting $F_{\epsilon}$ {\em does} commute with $\bf g$?
Of course, we can not, there are obstacles. 

\vspace{10pt}
\noindent
In this paper we will introduce a geometrical framework for describing these obstacles
and to what they correspond in the strong coupling limit {\it via} AdS/CFT duality.

\subsection{Holographic renormalization}

AdS/CFT correspondence relates the deformations of CFT to the classical solutions of SUGRA deforming AdS.
As main example, 
consider Type IIB SUGRA in $AdS_5\times S^5$ and $N=4$ SYM on the boundary $\partial(AdS_5\times S^5)$.
Deformations of the SYM action of the form (\ref{Deformations}) are mapped by AdS/CFT to the
classical SUGRA solutions, deformations of $AdS_5\times S^5$. Linearized SUGRA solutions correspond to
linearized deformations.

Renormalization of the deformations of QFT (Section \ref{sec:IntroRenorm}) should correspond
to {\em something} on the AdS side. Most of the work on {\em holographic renormalization} was done
along the lines of \cite{Bianchi:2001kw,Skenderis:2002wp}, and was based on the study of the bulk supergravity
action.

On the other hand, the computation of the renormgroup flow of the beta-deformation 
done in \cite{Aharony:2002hx} seems to use a different method. In particular, the authors of \cite{Aharony:2002hx}
did not need to know the action of the bulk theory. This, in particular, may allow to apply their
method to the cases where the action is not known and maybe even does not exists,
such as higher spin theories \cite{Vasiliev:2004qz,Sharapov:2017yde}.

\subsection{Geometrical abstraction}\label{sec:GeometricaAbstraction}
Suppose that a Lie algebra $\bf g$ acts on a manifold $M$, preserving a point $p$.
Then it acts in the tangent space to $M$ at $p$. The question is, can we find a formal map:
\begin{align}\label{IntroMapF}
  F_{\epsilon}\;:\;T_pM \rightarrow M
  \\
  F_0 \equiv p
\end{align}
parametrized by $\epsilon$
(``formal'' means power series in $\epsilon$) from the tangent space to $m$ to $M$, commuting with
the action of $\bf g$?  There are, generally speaking, obstructions to the existence of such a map
--- see Section \ref{sec:Linearization}.
We want to classify these obstructions. This is, essentially, equivalent to studying the {\bf normal form
of the action} of $\bf g$ in the vicinity of the fixed point.

\paragraph     {Tangent vectors as equivalence classes of trajectories}
Maps $F_{\epsilon}\;:\;T_pM \rightarrow M$ participate in the ``usual'' definition of the tangent space
({\it e.g.} \cite{Arnold}).
The tangent space $T_pM$ is defined as the space of equivalence classes of paths (maps from $\bf R$ to $M$)  $p(\epsilon)$ such
that $p(0)=p$. The equivalence relation is that two paths $p_1$ and $p_2$ are equivalent when
$p_1(\epsilon)-p_2(\epsilon) = o(\epsilon)$ in a coordinate patch. Giving a function $F_{\epsilon}$ as in
Eq. (\ref{IntroMapF}) is same as giving a prescription of how to pick, for each tangent vector $v$,
one path from the corresponding equivalence class. That path is:
\begin{align}
p(\epsilon) = F_{\epsilon}(v)
\end{align}
Of course, there are many such prescriptions. The question is, can we find some, which would be consistent
with the action of $\bf g$?

The space of formal paths $p\;:\;{\bf R}\rightarrow M$ such that $p(0)=p$ can be denoted $\Omega_pM$
--- similar to the space of $p$-based loops in $M$, but we only need a formal power series
in $\epsilon$ at $\epsilon=0$, not the whole loop. To summarize, we investigate the existence of a map:
\begin{equation}\label{LiftToFormalLoops}
T_pM \rightarrow \Omega_pM
\end{equation}
commuting with the action of $\bf g$.
We find that there are obstacles to the existence of such a map, and classify them. These obstacles
are, roughly speaking, some cohomology groups. More precisely, they are solutions of a Maurer-Cartan
equation, modulo gauge transformations (a nonlinear analogue of cohomology groups) ---
see Section \ref{sec:Linearization}.

\paragraph     {The role of supergeometry and infinite-dimensional geometry}

In our main application (AdS/CFT): ${\bf g} = {\bf psu}(2,2|4)$ --- the superconformal algebra.
Its even (bosonic) part is $so(2,4)\oplus so(6)$. If $M$ were a finite-dimensional ``usual'' (not super)  manifold
then there would be no obstacle in linearizing the action, because the relevant cohomology groups are zero.
This makes our picture somewhat counter-unintuitive geometrically.

\commentstarts{\small
  The relevant cohomology group is $H^1$.
  In classical geometry, we would have nontrivial invariants if $\bf g$ were ${\bf u}(1)$ or contained
  ${\bf u}(1)$ as a subalgebra. For example, a classical mechanical system can have a free limit
  and have in that limit periodic trajectories, but away from that free limit the trajectories are not periodic. 
}\commentends

There are two reasons for having
nontrivial invariants. The first reason is that $M$ is actually infinite-dimensional. But there is
also the second reason: even when we can find some finite-dimensional ``subsectors'' (submanifolds in $M$),
they are actually {\em super}-manifolds. This can make the cohomolgy nontrivial even in finite-dimensional case.

\subsection{Summary of this paper}

\paragraph     {Cohomological framework for holographic renormalization}

Here we will develop a formalism for computations along the lines of \cite{Aharony:2002hx}, which
makes them geometrically transparent. We interpret \cite{Aharony:2002hx} as computing 
certain  {\em invariants of supergravity equations} in the vicinity of AdS,
namely the solution of some Maurer-Cartan equation modulo gauge transformations. 
We give the definition of these invariants in Section \ref{sec:GeometricaAbstractionAdS}.
This is broadly similar to the obstructions to the existence of the $\bf g$-invariant map $F_{\epsilon}$
of Section \ref{sec:GeometricaAbstraction}. The details, however, are more subtle, because we are dealing with gravity.
The symmetry algebra ${\bf g} = {\bf psu}(2,2|4)$ of $AdS_5\times S^5$ is
actually a part of the larger infinite-dimensional symmetry, the gauge symmetries of the supergravity theory.
This makes the analysis on the AdS side more interesting.

\paragraph     {Finite-dimensional representations have nontrivial cohomology}

The cohomological obstacles for linearization of the symmetry are usually rather complicated,
because they involve the cohomology with coefficients in infinite-dimensional representations
(see Section \ref{sec:InfiniteDimensionalExtension}).
But when the symmetry is {\em super}-symmetry, even finite-dimensional representations
have nontrivial cohomology (Section \ref{sec:ObstacleIsQuadratic}).
We formulate some conjecture\footnote{the computation required to prove or dispove this 
  is outlined in Section \ref{sec:OutlineOfComputation}}
about the role of these cohomology classes in the particular case of beta-deformation.
Our conjecture implies that
the anomalous dimension (which in the case of beta-deformation is cubic in the deformation parameter) is,
in some sense,
a square of a simlper obstruction, which is quadratic in the deformation parameter.
While the anomalous dimension is analogous to a four-point function, the simpler
quadratic obstruction is analogous to a three-point function. This might explain
the observation in \cite{Aharony:2002hx} that the anomalous dimension is not renormalized.

\vspace{10pt}
\noindent
The idea is, therefore, to study the space of {\em all} perturbative solutions of supergravity (instead of
particular solutions) and describe its invariants, as a invariants of a supermanifold.

\paragraph     {Plan}
In Section \ref{sec:Linearization} we develop geometrical formalism for studying
the obstructions to the existence of the map (\ref{IntroMapF}) commuting with $\bf g$.
We explain that the obstacle is a solution of the Maurer-Cartan (MC) equation with values in
vector fields. 
In Section \ref{sec:FeynmanDiagrams} we explain how to apply this formalism to the
space of perturbative solutions of a classical field theory. We show that there is
a natural operation of ``amputation of the last leg'' which converts Feynman
diagrams into a solution of the MC equation.
In particular, in Section \ref{sec:ClassicalCFT}
we consider the case of classical CFT in ${\bf R}\times S^{d-1}$.
In Section \ref{sec:AdS} we discuss deformations of $AdS_5\times S^5$ and holographic renormalization.
In Section \ref{sec:BetaDeformation} we study the particular case of beta-deformation \cite{Leigh:1995ep}.
Finally, in Section \ref{sec:Discussion} we discuss some open questions and
potential problems.

\section{Obstacles to linearization of symmetry}\label{sec:Linearization}
\subsection{Action of a symmetry in local coordinates}\label{sec:ActionOfSymmetry}
Suppose that a Lie algebra $\bf g$ acts on a manifold $M$ and leaves invariant 
a point $p\in M$. Then $\bf g$ naturally acts in the tangent space $T_pM$.
Consider maps $F_{\epsilon}\;:\;T_pM\rightarrow M$ parameterized by a small parameter $\epsilon$,
satisfying:
\begin{align}
  & F_{\epsilon}(0) = p
    \label{MustPassThroughP}\\   
  & F_{\epsilon *}(0) = {\bf 1}\;:\;T_pM \rightarrow T_pM
    \label{DerivativeIsId}\\   
  & F_{\kappa\epsilon}(x) = F_{\epsilon}(\kappa x)
    \label{RescaleEpsilon}
\end{align}
As we discussed in Section \ref{sec:GeometricaAbstraction}, there are many such maps. 
Let us ask the following question: is it possible to construct such a map $F_{\epsilon}$
which would also commute with the  action of $\bf g$?
(We are interested in a {\em formal} map, {\it i.e.} a map specified as an 
infinite series in $\epsilon$; we will not discuss convergence.)

Let us start by picking {\em some} map $F\;:\; T_pM\to M$ (not necessarily ${\bf g}$-invariant)
satisfying Eqs. (\ref{MustPassThroughP}), (\ref{DerivativeIsId}) and (\ref{RescaleEpsilon}).
For each element $\xi\in {\bf g}$ there is a corresponding vector field $v\langle\xi\rangle$ on $M$. Let us 
consider $F_{\epsilon *}^{-1}v\langle\xi\rangle$. It is a vector field on $T_pM$:
\begin{equation}\label{ExpansionOfVectorField}
   F_{\epsilon *}^{-1} v\langle\xi\rangle =
   v_0\langle\xi\rangle + \epsilon v_1\langle\xi\rangle + \epsilon^2 v_2\langle\xi\rangle + \ldots
\end{equation}
where $v_n\langle\xi\rangle$ is of the form $v_n\langle\xi\rangle=f_n^{\mu}(x){\partial\over\partial x^{\mu}}$ with $f_n^{\mu}(x)$ a polynomial of the degree $n+1$ in $x$.

Notice that the power of $\epsilon$ in Eq. (\ref{ExpansionOfVectorField}) correlates with the degree in $x$ of
$f_n^{\mu}(x)$. Therefore we will  just skip $\epsilon$ from our formulas; we will think of ``$x$ being of the
order $\epsilon$''. 

The vector field $F^{-1}_*v\langle\xi\rangle$ has a very straightforward meaning. Our map $F$ turns a sufficiently small
open neighborhood of $0\in T_pM$ into a chart of $M$. In this context, $F^{-1}_*v\langle\xi\rangle$ is just the
``coordinate representation'' of the vector field $v\langle\xi\rangle$ in that chart. Therefore, our question
becomes:
\begin{itemize}
\item Can we choose a chart so that $v_1=v_2=\ldots = 0$?
\end{itemize}
We will see that obstacles are certain cohomology classes.

\subsection{Maurer-Cartan equation}
For two elements $\xi$ and $\eta$ of $\bf g$, we have:
\begin{equation}
   \left[\;F_*^{-1} v\langle\xi\rangle\;,\;F_*^{-1} v\langle\eta\rangle\;\right] 
   \;=\; F_*^{-1} v\langle[\xi,\eta]\rangle
\end{equation}
This means that, for $c\in \Pi {\bf g}$:
\begin{equation}\label{CommutatorOfVectorFieldsDependingOnGhosts}
   \left[F_*^{-1} v\langle c\rangle \,,\, F_*^{-1} v\langle c\rangle\right] =
   c^Ac^Bf_{AB}{}^C {\partial\over\partial c^C}F_*^{-1} v\langle c\rangle
\end{equation}
where $c^A$, $A\in \{1,2,\ldots,\mbox{dim}({\bf g})\}$ denote the coordinates on $\Pi {\bf g}$. Besides that:
\begin{equation}
   [v_0\langle\xi\rangle,v_0\langle\eta\rangle] = v_0\langle[\xi,\eta]\rangle
\end{equation}
Define the ``BRST operator'':
\begin{equation}\label{BRSTOperator}
   Q = {1\over 2}c^Ac^Bf_{AB}{}^C {\partial\over\partial c^C} + v_0\langle c\rangle
\end{equation}
where $c = c^At_A\in {\bf g}$. We have $Q^2=0$. This defines the differential in the Lia algebra
cohomology complex \cite{Knapp} of $\bf g$ with values in the space of vector fields on $T_pM$ having zero
of at least second order at the point $p$. (The action of the second term, $v_0\langle c\rangle$,
is by the commutator of vector fields.)

Let us define $\Psi$ as follows ({\it cf.} Eq. (\ref{ExpansionOfVectorField})):\marginpar{$\Psi$}
\begin{equation}\label{DefPsi}
   \Psi = F_*^{-1}v\langle c\rangle - v_0\langle c\rangle =  v_1\langle c\rangle + v_2\langle c\rangle +\ldots
\end{equation}
Eq. (\ref{CommutatorOfVectorFieldsDependingOnGhosts}) implies that $\Psi$ satisfies the MC equation:
\begin{equation}\label{MaurerCartan}
   Q\Psi  + {1\over 2}[\Psi,\Psi] = 0
\end{equation}

\subsection{Gauge transformations}\label{sec:GaugeTransformations}
Suppose that we replace $F\;:\;T_pM\to M$ with another function $\widetilde{F}=F\circ G$, where
$G$ is any (nonlinear) function $T_pM\to T_pM$ such that $G(0)=0$ and $G_*(0)=\mbox{id}$. Then $\Psi$ gets replaced with
$\widetilde{\Psi}$ where:
\begin{equation}\label{GaugeTransformations}
   \widetilde{\Psi} = G_*^{-1}\Psi  + G_*^{-1}QG
\end{equation}
This is the gauge transformation. An infinitesimal gauge transformation is:
\begin{equation}\label{InfinitesimalGaugeTransformations}
   \delta_{\Phi}\Psi = Q\Phi + [\Psi,\Phi]
\end{equation}

\subsection{Tangent space to the moduli space of solutions of MC equation}

The tangent space to the solutions of Eq. (\ref{MaurerCartan}) at the point $\Psi$ is the cohomology
of the operator $Q + [\Psi,\_]$:
\begin{equation}
   T_{\Psi}(MC) = H^1(Q + [\Psi,\_])
\end{equation}
The Lie algebra of nonlinear\footnote{{\it i.e.} quadratic and higher orders in coordinates} vector fields
has a filtration by the scaling degree. Therefore the cohomology of $H^1(Q + [\Psi,\_])$ can be computed by a spectral sequence.
The first page of this spectral sequence is:
\begin{equation}\label{E1forTMC}
   E_1^{p,q} = H^q(Q, \mbox{Hom}(S^pL,L))
\end{equation}
where $L = T_pM$.

\subsection{Monodromy transformation}\label{sec:MonodromyTransformation}
\paragraph     {Additional assumption}
We now have two actions of $\bf g$ on $T_pM$: the linearized one, which is given by $v_0$ of
Eq. (\ref{ExpansionOfVectorField}), and the nonlinear action given by $F_*^{-1}v\langle\xi\rangle$.
Suppose that the linearized one integrates to some action $\rho_0$ of a group $G$:
\begin{align}
  & \rho_0\;:\; G \longrightarrow
    \mbox{Hom}(T_pM, T_pM)
  \\
  & \rho_{0*}(\xi) = \left.{d\over dt}\right|_{t=0}\rho_0\left(e^{t\xi}\right) \;=\; v_0\langle\xi\rangle
\end{align}
Suppose that this group $G$ has a non-contractible one-dimensional cycle.

Consider the path ordered exponential over this non-contractible cycle. Without loss of generality,
we can start and end the loop at the unit. We define:
\begin{align}\label{MonodromyTransformation}
  m \;=\;& P\exp\left[\oint \rho_0(g)^{-1}\Psi\langle dgg^{-1}\rangle \rho_0(g)\right] =
  \\  
   \;=\;& \left(P\exp\left[\oint \rho_0(dg g^{-1}) + \Psi\langle dgg^{-1}\rangle \right]\right)
\end{align}
where $\rho_0(g)$ is the action of $G$ on $T_0M$. This is the monodromy transformation:
\begin{equation}
   m\in \mbox{Map}(T_pM,T_pM)
\end{equation}
Notice that the derivative of $m$ at the point $0\in T_pM$ is zero. Therefore we can define
its second derivative:
\begin{equation}
m'' \in \mbox{Hom}(S^2T_pM, T_pM)
\end{equation}
Eq. (\ref{MonodromyTransformation}) implies:
\begin{align}
  m'' =
  & \oint \rho_0(g)^{-1}\Psi_2\langle dgg^{-1}\rangle \rho_0(g)
  \label{SecondDerivativeOfM}\\
  \mbox{\tt\small where }
  & \Psi_2 = v_1 \mbox{ \tt\small of Eq. (\ref{ExpansionOfVectorField})}
\end{align}
Usually the cycle is such that $\dot{g}g^{-1}$ is constant. Then the meaning of
the integration in Eq. (\ref{SecondDerivativeOfM}), is that that we pick the {\em resonant terms}
in the quadratic vector field $v_1$. 

\subsection{Symmetries}\label{sec:Symmetries}
The monodromy transformation $m$ commutes with the action of $\bf g$, but we have to remember that the action of
$\bf g$ is given by nonlinear vector fields --- see Eq. (\ref{ExpansionOfVectorField}).
If it were given just by linearized vector fields, {\it i.e.} the $v_0\langle \xi\rangle$ of Eq. (\ref{ExpansionOfVectorField}), life would be easier. But this is, generally speaking, not the case. Notice that
$v_0\langle\xi\rangle = \xi{\partial\over\partial c}$. Instead of $\left[\xi{\partial\over\partial c}Q\,,\,\Psi\right]$ being zero, we have:
\begin{equation}\label{Symmetries}
   \left[\xi{\partial\over\partial c}Q\,,\,\Psi\right]\;=\;
   \left[Q+\Psi\,,\,\xi{\partial\over\partial c}\Psi\right]
\end{equation}
But $m''$ of Eq. (\ref{SecondDerivativeOfM}) {\em does} commute with the undeformed action of $\bf g$ on $T_pM$
({\it i.e.} with $v_0$). This is because $v_{\geq 1}$ are of quadratic and higher order, and the first
derivative of $m$ vanishes.

Sometimes $m''$ is zero on some subspace $L\subset T_pM$. Then, on this subspace, we can define the
third derivative $m'''$. Suppose that,  in addition, the restriction of $v_1$ on $L$ is parallel to $L$, {\it i.e.}:
\begin{align}
  v_1'' \;:\;
  & S^2 T_pM \rightarrow T_pM
  \\
  \mbox{\tt\small is such that: }
  & v_1''(S^2L)\subset L
\end{align}
Then $m'''$ commutes with the undeformed action of $\bf g$ on $T_pM$. 

\subsection{Closed subsectors}\label{sec:ClosedSubsectors}
Suppose that $T_pM$, as a representation of $\bf g$, has an invariant subspace:
\begin{equation}
V \subset T_pM
\end{equation}
It may happen that the restriction of $\Psi$ to $V$ is tangent to $V$. This, essentially,
means that $F(V)$ is closed under the action of the symmetry. In particular, the monodromy
transformation of Section \ref{sec:MonodromyTransformation} acts within $V$.
The sufficient condition for this is:
\begin{equation}
   H^1\left(
      {\bf g}\;,\; \mbox{Hom}(S^nV\,,\, T_pM/V)
   \right) = 0
\end{equation}

\section{Relation to tree level Feynman diagrams}\label{sec:FeynmanDiagrams}

Here we will apply the formalism of Section \ref{sec:Linearization} to the case when
$M$ is the space of solutions of some classical nonlinear field equations, constructed
as perturbation of some zero solution $p\in M$. This is 
different from the context of deformations of QFT (Section \ref{sec:IntroRenorm}),
but the AdS/CFT correspondence establishes a relation between these two contexts (Section \ref{sec:AdS}).

\subsection{Perturbation theory as a map $TM\rightarrow \Omega M$}
Let us take $M$ to be the space of perturbative solutions $\phi$ of nonlinear equations of the form:
\begin{equation}\label{NonlinearEquation}
   L\phi = f(\phi)
\end{equation}
where $L$ is some linear differential operator, and $f(\phi)$ is a nonlinear function 
describing the interaction. We assume that $f$ is a polynomial starting with 
the terms of quadratic or higher order.

The point $p\in M$ will be the zero solution $p=0$.
Then $T_0M$ can be identified with the space of solutions of 
the linearized equation:
\begin{equation}\label{LinearizedEqM}
   L\phi = 0
\end{equation}
Tree level perturbation theory can be thought of as a 1-parameter map
\begin{equation}
F_{\epsilon}\;:\; T_0M \rightarrow M
\end{equation}
parameterized by a small parameter $\epsilon$. As explained in Section \ref{sec:GeometricaAbstraction},
it can be also understood as a map $T_0M \rightarrow \Omega_0M$.

We will embed $M$ into the space $M_{\rm os}$ of all  field configurations,
not necessarily satisfying equations of motion (subindex ``os'' means ``off-shell``).
We assume that $M_{\rm os}$ is a linear space. We consider $F\;:\;T_0M \rightarrow M$
as a function $T_0M \rightarrow M_{\rm os}$. It can be described as a sum of tree level Feynman 
diagrams. Every incoming leg corresponds to a solution of the linearized equation 
(\ref{LinearizedEqM}). Every internal leg and the outgoing leg each correspond to a propagator $L^{-1}$. 
There is a recursion relation\footnote{the right hand side is a sum of two elements of $M_{\rm os}$; remember
that we assumed that $M_{\rm os}$ a linear space}:
\begin{equation}\label{RecursionRelation}
   F[\phi_0] = \epsilon\phi_0\;+ L^{-1} f(F[\phi_0])
\end{equation}
where $L^{-1}$ satisfies:
\begin{equation}\label{LInverse}
   LL^{-1} = {\bf 1}
\end{equation}
The definitions of the operator $L^{-1}$  has an ambiguity
(because one can add a solution of the free equation). Suppose that we made some choice of $L^{-1}$.
The dependence on the choice of $L^{-1}$
is controlled by Lemma \ref{theorem:GaugeTransformation} below.

As we already explained, we need an embedding of $M$ into the linear space of
off-shell field configurations $M_{\rm os}$, just because we want to add Feynman diagrams.
Obvously, the space $T_0M$ of free solutions is also embedded into $M_{\rm os}$. 
Let us assume that the action of $\bf g$ on $M_{\rm os}$ agrees with this embedding.
This is not really important, but we make this assumption for this Section.
For example, suppose $\bf g$ contains time translation $\partial\over\partial t$.
We assume that it acts as $\delta \phi = \partial_t \phi$, both on $M$ and on $T_0M$.

Let us define $\Psi$ as follows:
\begin{align}
  \Psi\;\in\;
  & \mbox{Hom}\left(\Pi {\bf g}\;,\; \mbox{Vect}(T_0M)\right)
  \\
  \Psi\langle c\rangle[\phi_0] \;=\;
  & [Q,L^{-1}]f(F[\phi_0])
    \label{PsiViaFeynmanDiagrams}
\end{align}
(the dependence of $c$ on the RHS comes from $Q$).

\paragraph     {Lemma \arabic{Theorems}:\label{theorem:SamePsi}}\refstepcounter{Theorems}
This $\Psi$ is the same $\Psi$ as defined in Eq. (\ref{DefPsi}): 

\begin{equation}
   \Psi = F_*^{-1} v\langle c\rangle - v_0\langle c\rangle
\end{equation}
\paragraph     {Proof} We have to show that for any  $F_*(v_0\langle c\rangle + \Psi) = v\langle c \rangle$.
In other words, for any $\xi\in {\bf g}$:
\begin{align}\label{FStarOnRecurrentRelation}
   F_*\left(v_0\langle \xi\rangle + [\xi,L^{-1}] f(F[\phi_0])\right) = \xi F[\phi_0]
\end{align}
We will use:
\begin{equation}
   F[\phi_0]_* \;=\; {\bf 1} + L^{-1} f_* F[\phi_0]_*
   \quad \mbox{\tt\small therefore}\quad
   F[\phi_0]_*^{-1} \;=\; {\bf 1} - L^{-1}f_*
   \label{InverseDerivativeMap}
\end{equation}
We have:
\begin{align}
  \xi F[\phi_0] = \xi\phi_0 + [\xi, L^{-1}] f(F[\phi_0]) + L^{-1} f_*\xi F[\phi_0]
\end{align}
Together with Eq. (\ref{InverseDerivativeMap}) this implies:
\begin{align}
F[\phi_0]_*^{-1}\xi F[\phi_0]\;=\;\xi\phi_0 + [\xi, L^{-1}] f(F[\phi_0])
\end{align}
The proof can be put in slightly different words, as follows. Notice:
\begin{align}
   QF[\phi_0] \;&= Q\left( 
      \phi_0 + L^{-1}f(\phi_0) + L^{-1}f(L^{-1}f(\phi_0)) + \ldots
   \right)
\end{align}
Every time $Q$ hits $\phi_0$, we get $v_0\langle c\rangle$:
\begin{equation}
   \left.{d\over dt}\right|_{t=0}\Big(
      tv_0\langle c\rangle
      + L^{-1}f(\phi_0 + tv_0\langle c\rangle)
      + L^{-1}f(L^{-1}f(\phi_0 + tv_0\langle c\rangle)) 
      + \ldots
   \Big)
\end{equation}
--- this gives the $F_*v_0\langle c\rangle$ term on the LHS of Eq. (\ref{FStarOnRecurrentRelation}). And when $Q$ hits one of the $L^{-1}$,
we get $F_*\Big([Q,L^{-1}]f(F[\phi_0])\Big)$.

\paragraph     {Lemma \arabic{Theorems}:\label{theorem:GaugeTransformation}}\refstepcounter{Theorems}
An infinitesimal variation of $L^{-1}$:
\begin{align}
  & L^{-1} \mapsto L^{-1} + \delta L^{-1}
\end{align}
where $\delta L^{-1}$ satisfies $L \delta L^{-1} = 0$,
corresponds to an infinitesimal gauge transformation of $\Psi$ (see Eq. (\ref{InfinitesimalGaugeTransformations})) where:
\begin{equation}
\Phi = \delta L^{-1} f(F[\phi_0])
\end{equation}

\paragraph     {Proof}
\begin{align}
  \delta\Psi(\phi_0) \;=\;
  &  \delta \left([Q,L^{-1}] f(F[\phi_0])\right) =
  \nonumber\\   
  \;=\;
  & [Q,\delta L^{-1}] f(F[\phi_0]) + [Q,L^{-1}] f(F[\phi_0])_* \delta F[\phi_0]\;=
  \nonumber\\   
  \;=\;
  & [Q,\Phi] - \delta L^{-1} f(F[\phi_0])_* F[\phi_0]_* \Psi + [Q,L^{-1}] f(F[\phi_0])_* \delta F[\phi_0]\;=
  \nonumber\\  
  \;=\;
  & [Q,\Phi] + [\Psi,\Phi]
\end{align}

\subsection{Amputation of the last leg}\label{sec:AmputationOfTheLastLeg}
We will now present a slightly different point of view on the construction.
Suppose that for every linearized solution $\phi_0$ we  constructed a nonlinear solutions $\phi$
(depending on a small parameter $\epsilon$). What should we do with $\phi$, to obtain $\Psi\langle c\rangle$?
Remember that $\Psi\langle c\rangle$ is a (nonlinear) vector field on the space of
linearized solutions. Obviously, we have to somehow ``project'' $\phi$ to a linearized solution.
According to Eq. (\ref{PsiViaFeynmanDiagrams}) we should remove the last leg, and replace it
with $[Q,L^{-1}]$:
\begin{equation}
   \Psi = [Q,L^{-1}]f(\phi) = [Q,L^{-1}]L \phi
\end{equation}
Remember that $L^{-1}$ satisfies Eq. (\ref{LInverse}):
\begin{equation}
LL^{-1}= {\bf 1}
\end{equation}
Let us define the ``amputator'' $A$ as the composition:
\begin{align}
  A := [Q,L^{-1}]L\;=\;& [P,Q]
  \\
  \mbox{\tt\small where } & P = \left({\bf 1} - L^{-1}L\right)
\end{align}
(Notice that $P$ is a projector to $\mbox{ker}L$.) 
It satisfies\footnote{Actually, any operator of the form $[P,Q]$, where $P^2=P$ and $PQP=QP$, 
is nilpotent; the nilpotence of $Q$ is not necessary for the nilpotence of $A$.}:
\begin{equation}
A^2 = 0
\end{equation}
If $\phi$ is our perturbative solution ({\it i.e.} $\phi = \phi_0 + L^{-1} f(\phi)$), then:
\begin{equation}
[Q,L^{-1}]f(\phi) = A\phi
\end{equation}
This leads to the following interpretation. The ``projector'' $P$ can be interpreted as a map $M\rightarrow T_0M$
(Section \ref{sec:ActionOfSymmetry}), the inverse of $F$. Then, again, $[P,Q] = F_*^{-1}v - v_0$.

\subsection{Trivial example}\label{sec:TrivialExample}

Consider a vector field ${\bf V}\in \mbox{Vect}({\bf R}^n)$.
Suppose that $\bf V$ vanishes at $0\in {\bf R}^n$, and the derivative of $\bf V$ also vanishes at $0$:
\begin{equation}
   {\bf V}(0) = 0 \mbox{ \tt\small and } {\bf V}'(0)=0
\end{equation}
Consider the following equation:
\begin{equation}\label{FlowEquation}
   {d\over dt} {\bf x}(t) = {\bf V}({\bf x}(t))
\end{equation}
In notations of Section \ref{sec:FeynmanDiagrams}, $M$ be the space of all solutions of Eq. (\ref{FlowEquation})
and $L = d/dt$.
This equation is invariant under translations of $t$, generating ${\bf g} = {\bf R}$.
The generator of ${\bf g}$ is $\xi = d/dt$. 

Let us construct solutions perturbatively in the vicinity of the constant solution:
\begin{align}
  p\;:\; & {\bf x}(t) = 0
  \\
  T_0M \;=\;& {\bf R}^n \mbox{ \tt\small (constant $\bf x$)}
\end{align}
For any functions $f(t)$, we can expand it in Taylor series around $t=0$. We define $L^{-1}$ as follows:
\begin{equation}
   L^{-1} t^m = {1\over  m+1} t^{m+1}
\end{equation}
The map $F_{\epsilon}\;:\;T_0M \rightarrow M$ of Eq. (\ref{RecursionRelation}) is:
\begin{equation}
   {\bf x}_0 \mapsto \epsilon {\bf x}_0 + t {\bf V}(\epsilon {\bf x}_0) + \ldots
\end{equation}
We observe:
\begin{equation}\label{AmputatorTrivial}
   \left[ {d\over dt} \, , \, L^{-1} \right] t^m
   \;=\;
   \left\{\begin{array}{l}
            1 \mbox{ \tt\small if } m=0
            \cr
            0 \mbox{ \tt\small if } m>0
          \end{array}
       \right. 
    \end{equation}
    According to Eq. (\ref{PsiViaFeynmanDiagrams}), to construct $\Psi$ we have to
    take ${\bf V}(\epsilon {\bf x}_0 + t {\bf V}(\epsilon {\bf x}_0) + \ldots)$ and hit it with
    $\left[ {d\over dt} \, , \, L^{-1} \right]$, which, according to Eq. (\ref{AmputatorTrivial}) amounts
    to put $t=0$.
    Therefore, in this case:
    \begin{equation}
       \Psi = c {\bf V}
    \end{equation}
    The gauge transformations of Eq. (\ref{InfinitesimalGaugeTransformations}) are:
    \begin{equation}
       \delta_{\Phi} \Psi = c [{\bf V},\Phi]
    \end{equation}
    where $\Phi\in\mbox{Vect}({\bf R}^n)$ is another vector field.
    Therefore, in this case the MC invariant computes the normal form of the vector field
    $\bf V$ ({\it i.e.} $\bf V$ modulo nonlinear changes of coordinates).

    In this example we had ${\bf g} = {\bf R}$, and the structure of the first cohomology
    group was rather tautological: $H^1({\bf g}, L)$ was the space of $\bf g$-invariants in $L$. 
    In the next example we will consider the non-abelian ${\bf g} = so(2,d)$,
    with much more interesting (smaller!) cohomologies.
    In particular, $H^1({\bf g},V)$ can be nonzero only for infinite-dimensional $V$.

\subsection{Example: classical CFT on ${\bf R}\times S^{d-1}$}\label{sec:ClassicalCFT}

Consider a conformally invariant classical theory on the Lorentzian ${\bf R}\times S^{d-1}$, for example
the $\phi^4$ theory, $d=4$.

\subsubsection{Realization of ${\bf R}\times S^{d-1}$ as the base of the lightcone}
We will use same notations as in Appendix \ref{sec:EmbeddingFormalism}. We denote:
\begin{equation}
d = D - 1
\end{equation}
Consider the light cone in ${\bf R}^{2,D-1}$ ({\it cp} Eq. (\ref{Hyperboloid})):
\begin{equation}\label{LightCone}
   I^2 := |Z|^2 - \sum\limits_{i=1}^d X_i^2 = 0
\end{equation}
A convenient model for the conformal ${\bf R}\times S^{d-1}$ is the projectivization of the light
cone, which is parametrized by $(Z,X_1,\ldots,X_d)$ satisfying (\ref{LightCone}) modulo the 
equivalence relation:
\begin{equation}
   (Z,X_1,\ldots,X_d) \simeq (\lambda Z,\lambda X_1,\ldots,\lambda X_d)\;,\;\;
   \lambda\in{\bf R}
\end{equation}
A density of the weight $w$ is a function $\sigma(Z,X_1,\ldots,X_d)$ satisfying:
\begin{equation}
   \sigma(\lambda Z,\lambda X_1,\ldots,\lambda X_d) = \lambda^{-w} \sigma(Z,X_1,\ldots,X_d)
\end{equation}
modulo functions divisible by $I^2$. Let ${\bf D}_w$ denote the space of such densities. The
conformally invariant d'Alambert acts as follows:
\begin{align}
   L \;:&\; {\bf D}_{d-2\over 2} \rightarrow {\bf D}_{d+2\over 2}
   \\
   L\;\sigma \;=&\; 
   \left(
      4{\partial\over\partial Z}{\partial\over\partial \overline{Z}} - 
      \sum\limits_{i=1}^d {\partial^2\over \partial X_i^2}
   \right)\sigma
\end{align}
This operator is only well-defined with this value of $w$, because for other values
of $w$ it would not annihilate modulo $I^2$ those functions which are divisible by $I^2$.

The elements of the kernel of $L$, {\it i.e.} the solutions of free field equations, are
real sums of positive and negative frequency waves:
\begin{align}
  \sigma \;=\;
  & \sigma_+ + \sigma_-
  \label{FreeSolutions}\\  
  \mbox{\tt\small where }
  & \sigma_+ \;=\; {p(X)\over Z^{\mbox{deg}(p) + {d-2\over 2}}}
  \\  
  & \sigma_- \;=\; {\overline{p}(X)\over \overline{Z}^{\mbox{deg}(\overline{p}) + {d-2\over 2}}}
\end{align}

\subsubsection{Conformal symmetry}
Besides the rotations of $S^{d-1}$, there are also the following conformal 
transformations:
\begin{align}
  E   \;& = Z{\partial\over\partial Z} - \overline{Z}{\partial\over\partial\overline{Z}}
  \label{GenE}\\    
  K_i \;& = 2 X_i {\partial\over\partial Z} +  \overline{Z} {\partial\over\partial X_i}
  \label{GenK}\\    
  \overline{K}_i \;
        & = 2 X_i {\partial\over\partial \overline{Z}} +  Z {\partial\over\partial X_i}
        \label{GenBarK}             
\end{align}

\subsubsection{Amputator}\label{sec:Amputator}
Introduce the Lie algebra cocycle $C$: 
\begin{align}
  C \;\in\;
  & Z^1\left(
    {\bf g}\;,\;
    \mbox{Hom}\left(
    {\bf D}_{d/2 + 1}\,,\,
    \mbox{ker}\left({\bf D}_{d/2 - 1}\stackrel{L}{\longrightarrow} {\bf D}_{d/2 + 1}\right)
    \right)
    \right)
    \nonumber \\
  C\;=\;
  & [Q, L^{-1}]
  \label{CocycleC} 
\end{align}
As we explained in Section \ref{sec:AmputationOfTheLastLeg}, given a perturbative solution $F[\phi_0]$,
the corresponding solution of the MC Eq. (\ref{MaurerCartan}) is:
\begin{equation}
\Psi = Cf(F[\phi_0])
\end{equation}
Consider elements of  ${\bf D}_{d+2\over 2}$  periodic in global time. Any such element can be written as:
\begin{equation}\label{PeriodicDensity}
 \sigma =  \sum_{\rho,\overline{\rho}} {p_{\rho,\overline{\rho}}(X)\over \overline{Z}^{\overline{\rho}} Z^{\rho}}
\end{equation}
where the summation is over a pair of integers $\rho,\overline{\rho}$ and $p_{\rho,\overline{\rho}}(X)$ is a 
{\em harmonic polynomial} of $X_1,\ldots,X_d$ of the following degree: 
\begin{equation}
\mbox{deg}(p_{\overline{\rho},\rho}) = \overline{\rho} + \rho - {d+2\over 2}
\end{equation}
We define $L^{-1}\;:\;{\bf D}_{d+2\over 2}\rightarrow {\bf D}_{d-2\over 2}$ as follows:
\begin{equation}
    L^{-1}\left({p(X)\over \overline{Z}^{\overline{\rho}} Z^{\rho}}\right) \;=\;
    \left\{ \begin{array}{rl} 
              \mbox{\tt if }\rho\neq 1 \mbox{ \tt and } \overline{\rho}\neq 1\;:\;
              & {1\over 4}{1\over (\overline{\rho} - 1)(\rho - 1)}\;{p(X)\over \overline{Z}^{\overline{\rho} - 1} Z^{\rho - 1}}
                \cr
                \mbox{\tt if }\rho = 1\;:\;
              & - {1\over 4}{1\over (\overline{\rho} - 1)}\left(\mbox{log} {Z\over\overline{Z}}\right)\;
                {p(X)\over \overline{Z}^{\overline{\rho} - 1}}
                \cr
                \mbox{\tt if }\overline{\rho} = 1\;:\;
              & {1\over 4}{1\over (\rho - 1)}\left(\mbox{log} {Z\over\overline{Z}}\right)\;
                {p(X)\over Z^{\rho - 1}}
                \cr
                \mbox{\tt if }\rho = \overline{\rho} = 1\;:\;
              & - {1\over 8}\left(\mbox{log} {Z\over\overline{Z}}\right)^2p(X)\;
    \end{array}
    \right.
\label{Propagator}
\end{equation}
Therefore:
\begin{align}
[L^{-1}, K_i]
\left( {p(X)\over \overline{Z}^{\overline{\rho}}Z} \right) \;& =\;
{1\over 4} {\partial_i p(X)\over (\overline{\rho}-1)^2\overline{Z}^{\overline{\rho}-2}}
\label{AK}\\      
[L^{-1},\overline{K}_i]
\left( {p(X)\over \overline{Z}^{\overline{\rho}}Z} \right) \;& =\;
{1\over 4}{\left[||X||^2 \partial_i - 2X_i (X_j \partial_j) - (d-2) X_i\right]\;p(X)
\over (\overline{\rho}-1)^2 \overline{Z}^{\overline{\rho}}}
\label{AKbar}\\ 
[L^{-1},E] 
\left({p(X)\over \overline{Z}^{\overline{\rho}} Z}\right)
\;& =\;
{1\over 2} {p(X)\over 
(\overline{\rho}-1) \overline{Z}^{\overline{\rho}-1}}
\label{AE}
\end{align}
These formulas {\em partially} define a cohomology class $C$ of Eq. (\ref{CocycleC}). The definition
is only partial, because Eq. (\ref{PeriodicDensity}) does not describe all elements of ${\bf D}_{d+2\over 2}$,
but only those periodic which are periodic in the global time $t= {1\over i}\mbox{log}{Z\over\overline{Z}}$.
Generic elements are linear combinations of:
\begin{equation}\label{DensityWithPowersOfT}
t^k {p(X)\over \overline{Z}^{\overline{\rho}} Z^{\rho}}
\end{equation}
To completely specify $C$, we have to define $L^{-1}$ on elements containing powers of $t$, and compute
for them the commutators, as in Eqs. (\ref{AK}), (\ref{AKbar}), (\ref{AE}). We will not do it here.

\subsubsection{Relation to renormgroup}
Our discussion of the classical field theory solutions in this section is a warm-up.
However, it {\em is} related to renormalization. Given a set of operators $\{{\cal O}_I\}$,
{\it e.g.} ${\cal O}_{i_1,\ldots i_N} = \partial_{i_1}\cdots\partial_{i_N}\phi$,
and a set of infinitesimal coefficients $\epsilon_I$, let us define the coherent state, schematically:
\begin{equation}\label{CoherentState}
\exp \left( \sum_I\epsilon_I{\cal O}_I\right)
\end{equation}
which in the classical limit corresponds to a classical solution.
This, of course, requires regularization. Therefore, the map
\begin{equation}\label{QuantumMap}
\sum_I\epsilon_I{\cal O}_I \mapsto \left.\exp \left( \sum_I\epsilon_I{\cal O}_I\right)\right|_{\rm renormalized}
\end{equation}
does not commute with the action of the symmetries. What we studied in this section
must be the classical limit of this map. This requires further study.

\subsection{Comments on the structure of $\Psi$}
\subsubsection{$\Psi$ is simpler than perturbative classical solutions}\label{sec:PsiIsSimple}
Let us continue with the example of the previous section.
Generally speaking, a perturbative solution is a sum of expressions of the form:
\begin{equation}
   t^k{p(X)\over \overline{Z}^{\overline{\rho}} Z^{\rho}}
   \quad \mbox{\tt\small where } \mbox{deg}(p) = \rho + \overline{\rho} - {d-2 \over 2}
\end{equation}
However, after the replacement of the last leg with $C$ of eq. (\ref{CocycleC}), the resulting expression
does not contain ``bare'' $t$ ({\it i.e.} only contains $t$ {\it via} its exponentials). 
Indeed, $Cf(\phi)$ is {\em a solution of the free field equations}. Solutions of the free field
equations do not contain powers of $t$. They only involve expressions of the form Eq. (\ref{FreeSolutions}).
No powers of $t$. In this sense, the amputated $\phi$ is much simpler than full perturbative solution.

\commentstarts{\small
  As we mentioned at the end of Section \ref{sec:Amputator}, we did not actually compute the
  amputator on the field configurations containing powers of $t$ (Eq. (\ref{DensityWithPowersOfT})).
  But we know in advance that the resulting expression will not contain any powers of $t$.
}\commentends

\vspace{5pt}
\noindent
Moreover, we know that $\Psi$ satisfies a constraint: the Maurer-Cartan Eq. (\ref{MaurerCartan}).
In some situations, this might allow for some partial bootstrap, see Section \ref{sec:Bootstrap}.

There is a price to pay: the definition of $\Psi$ contains an ambiguity. We could
have choosen a different $L^{-1}$. This corresponds to the gauge transformation of Eq. (\ref{GaugeTransformations}).
Moreover, the condition of $\bf g$-covariance is complicated:

\subsubsection{The condition of $\bf g$-covariance is complicated}
({\it cp} Section \ref{sec:Symmetries})

Under the false impression that all non-covariance is due to the ``resonant'' factors
$\mbox{log}{Z\over\overline{Z}}$, one might conjecture that $\Psi\langle\xi\rangle$ is ${\bf g}$-covariant in
the sense that:
\begin{equation}\label{FalseCovariance}
[\eta,\Psi\langle\xi\rangle] = \Psi\langle [\eta,\xi]\rangle \quad\mbox{(wrong)}
\end{equation}
This, however, is not the case. 
At least when $\bf g$ is a semisimple Lie algebra, Eq. (\ref{FalseCovariance}) is incompatible
with the MC equation:
\begin{equation}
   \Psi\langle [\eta,\xi] \rangle = [\eta,\Psi\langle\xi\rangle] - [\xi,\Psi\langle\eta\rangle] +
   [\Psi\langle\eta\rangle,\Psi\langle\xi\rangle] 
\end{equation}
because $\Psi$ takes values in vector fields of degree 1 and higher.
A semisimple Lie algebra cannot be represented by the vector fields
of degree 1 and higher. In fact, it follows immediately from Eqs. (\ref{AK}), (\ref{AKbar}) and (\ref{AE}) that with $\bf L$ chosen as in Eq. (\ref{Propagator})  $\Psi(\xi)$ is zero when $\xi$ is an infinitesimal rotation of $S^{d-1}$. This already contradicts Eq. (\ref{FalseCovariance}).

Instead of simple but wrong Eq. (\ref{FalseCovariance}) we have more complicated Eq. (\ref{Symmetries}).

\subsubsection{Can $\Psi$ be bootstrapped?}\label{sec:Bootstrap}
Consider an infinitesimal $G$-preserving deformation $s$ of the action which is a monomial
of the order $n$ in the elementary fields.
Then the corresponding cocycle,
representing a class of $H^1\left(Q,\mbox{Hom}(S^{n-1} T_0{\cal S},T_0{\cal S})\right)$, is given by the expression:
\begin{equation}\label{TermsOfLowestOrder}
[Q,L^{-1}] {\delta^{n-1}s\over \delta \phi^{n-1}}
\end{equation}
Could it be that {\em all}  $H^1\left(Q,\mbox{Hom}(S^{n-1} T_0{\cal S},T_0{\cal S})\right)$ is exhausted
by the expressions of this form for various ${\bf g}$-preserving deformations?
This is certainly {\em not} true for SUGRA on $AdS_5\times S^5$.
But in the situations when this is true, Eq. (\ref{MaurerCartan}) allows to recursively compute $\Psi$, modulo
gauge equivalence described in Section \ref{sec:GaugeTransformations},
starting from the terms of the lowest order in $\phi$ given by Eq. (\ref{TermsOfLowestOrder}).

\section{Supergravity in AdS space}\label{sec:AdS}

\subsection{Holographic renormgroup}

Consider Type IIB SUGRA in $AdS_{1,4}\times S^5$ and $N=4$ supersymmetric Yang-Mills on the boundary.
We can proceed in two ways, which are equivalent because of the AdS/CFT duality: 

\paragraph     {- Renormgroup flow on the boundary} We choose some map from the space
of linearized deformations of the $N=4$ SYM theory to the space of finite deformations.
There is no way to fix such a map preserving $\bf g$,
so we want to study the deviation from $\bf g$-invariance, in the context
of Section \ref{sec:Linearization}.

\paragraph     {- Classical solutions of SUGRA in the bulk} We fix some map
from the space of solutions of the linearized SUGRA equations to the space of
nonlinear solutions. Then we study the deviation of this map from being
$\bf g$-invariant as in Section \ref{sec:Linearization}.

\vspace{10pt}
\noindent
Here we will discuss this second approach. In fact, $\bf g$ is a subalgebra of a larger superalgebra,
the superalgebra of gauge transformations of supergravity.
This requires a generalization of the formalism of Section \ref{sec:Linearization}
which we will describe in Section \ref{sec:GeometricaAbstractionAdS}.

\subsection{Gauge transformations of supergravity}\label{sec:GaugeTransformationsSUGRA}

The precise description of the gauge transformations of supergravity depends, generally
speaking, on the formalism.
Any theory of supergravity necessarily includes the group of space-time super-diffeomorphisms
(= coordinate transformations), as gauge transformations.
Those theories which have B-field should also include gauge transformation of the B-field.
In the case of bosonic string they have been recently discussed in \cite{Schulgin:2014gra}.
In bosonic string, gauge transformations correspond to BRST exact vertices \cite{Schulgin:2014gra}.

In the pure spinor formalism, we are not aware of any reference discussing specifically
gauge transformations. Apriori the coordinate transformations should also include
transformations of pure spinor ghosts; they are in fact gauge-fixed \cite{Berkovits:2001ue}.

Our approach to holographic renormgroup is based on the study of the normal form of the action of the
group of SUGRA gauge transformations in the vicinity of the AdS solution.
We will now develop a geometrical abstraction for that.
The construction  parallels Section \ref{sec:GeometricaAbstraction}, the main difference being
that instead of a point $p\in M$ we have to consider a degenerate orbit ${\cal O}\subset M$.
We will now explain the details.

\subsection{Geometrical abstraction}\label{sec:GeometricaAbstractionAdS}

Consider a Lie supergroup $A$ acting on a supermanifold $M$ with a subgroup $G\subset A$
preserving a point
$p\in M$.
Moreover, we will assume that:
\begin{itemize}\item The action of $A$ on $M$ is free in a neighborhood of $p$ except at the orbit of $p$\end{itemize}
Let $\cal O$ denote the orbit of $p$:
\begin{equation}
   {\cal O} = Ap
\end{equation}
In the context of holographic renormgroup:
\begin{itemize}
\item $M$ is the space of SUGRA solutions in the vicinity of pure AdS
\item $p$ is $AdS_5\times S^5$ with some fixed choice of coordinates and zero $B$-field
\item $A$ is the group of all gauge transformations of supergravity, and ${\bf a} = \mbox{Lie} A$ is
   its Lie superalgebra
\item $G\subset A$ is the subgroup preserving $AdS_5\times S^5$, and ${\bf g} = {\bf psu}(2,2|4)$ its Lie superalgebra
\item $\cal O$ is the space of all metrics which can be obtained from the fixed $AdS_5\times S^5$ metric
   by coordinate redefinitions, and exact $B$-field
\end{itemize}
Introduce the coordinates $\alpha^I$ on $\cal O$, and coordinates $x^a$ in the transverse direction
to $\cal O$.
Consider the normal bundle $N{\cal O}$ of $\cal O$:
\begin{equation}
   0 \longrightarrow T{\cal O} \longrightarrow TM|_{\cal O} \longrightarrow N{\cal O} \longrightarrow 0
   \end{equation}
The action of $A$ on $M$ induces the action of $A$ on $N{\cal O}$. Let $I^{\infty} {\cal O}$
denotes the formal neighborhood of $\cal O$ in $M$. It also comes with an action of $A$.

Consider the following question: can we find a family of maps, parameterized by $\epsilon\in{\bf R}$:
\begin{equation}
   F_{\epsilon} \;:\; N{\cal O} \longrightarrow I^{\infty}{\cal O}
\end{equation}\label{MapFromNOToI}
satisfying ({\it cp} Eqs. (\ref{MustPassThroughP}), (\ref{DerivativeIsId}), (\ref{RescaleEpsilon})):
\begin{align}
  & F_0 \;=\; \pi : N{\cal O} \longrightarrow {\cal O}
    \label{MustPassThroughO}\\
  & \left.{d\over d\epsilon}\right|_{\epsilon = 0}f\circ F_{\epsilon}\circ \gamma \;=\;
    {\cal L}_{\gamma} f \quad \forall \gamma \in \Gamma(N{\cal O}), f\in \mbox{Fun}(M), f|_{\cal O} = 0,
    \label{DerivativeInNormalDirection}\\
  & F_{\kappa\epsilon} \;=\; F_{\epsilon}\circ R_{\kappa} \quad \mbox{where $R_{\kappa}$ rescales by $\kappa$ in the fiber}
    \label{RescaleEpsilonO}
\end{align}
commuting with the action of $A$?

It turns out that the obstacle exists already at the linearized level.
Normal bundle is not the same as the first infinitesimal neighborhood.
The space of functions on $I^1{\cal O}$ is not the same, as a representation of $\bf a$, as the
space of functions on $N{\cal O}$ constant-linear on fibers --- see Eq. \eqref{ObstacleToNI} below.

We will now proceed to the study of the obstacles to the existence of $F_{\epsilon}$.

\subsection{Normal form of the action of $\bf a$}\label{sec:NormalFormA}

Locally near the point $p$, the normal bundle to $\cal O$ can be trivialized. Let $\alpha^I$ denote
the coordinates on $\cal O$, and $x^a$ coordinates in the fiber. Then, the equation of $\cal O$ is:
\begin{equation}
   x^a = 0
\end{equation}

\subsubsection{General vector field tangent to $\cal O$}\label{GeneralVectorField}

Let us fix some $F_{\epsilon}$ satisfying Eqs. (\ref{MustPassThroughO}), (\ref{DerivativeInNormalDirection}),
(\ref{RescaleEpsilonO}).

Since $A$ acts on $M$, every element of the Lie algebra $\xi\in\bf a$ defines a vector field $v\langle\xi\rangle$ on $M$. 

Then, $F_{\epsilon *}^{-1} v\langle\xi\rangle$ is a vector field on $N{\cal O}$:
\begin{align}
  F_{\epsilon *}^{-1}v\langle\xi\rangle \;=\;
  &
    \phantom{+} \left(
    (u\langle\xi\rangle(\alpha))^I
    + \sum_{n\geq 1}\epsilon^n(\Theta\langle\xi\rangle(\alpha))^I_{b_1\cdots b_n} x^{b_1}\cdots x^{b_n}
    \right)
    {\partial\over\partial \alpha^I} \;+
  \nonumber \\\mbox{}
  &
    + \left(
    (v_0\langle\xi\rangle(\alpha))^a_b x^b
    + \sum_{n\geq 2}\epsilon^{n-1}(\Psi\langle\xi\rangle(\alpha))^a_{b_1\cdots b_n} x^{b_1}\cdots x^{b_n}
    \right)
    {\partial\over\partial x^a}
\end{align}
It should satisfy:
\begin{equation}\label{CommutatorInIO}
   [F_{\epsilon *}^{-1}v\langle\xi\rangle\,,\,F_{\epsilon *}^{-1}v\langle\eta\rangle]
   =
   F_{\epsilon *}^{-1}v\langle[\xi,\eta]\rangle
\end{equation}
The power of $\epsilon$ correlates with the degree in $x$-expansion. As an abbreviation, we will omit $\epsilon$
in the following formulas.

\paragraph     {Big Maurer-Cartan equation}

Let us introduce the Faddeev-Popov ghosts $\hat{c}\in \Pi {\bf a}$ for $\bf a$.
Let us  denote:
\begin{align}\hat{Q}\;=\;
 &
           (u\langle \hat{c}\rangle (\alpha))^I{\partial\over\partial\alpha^I}
           + (v_0\langle \hat{c}\rangle(\alpha))^a_b x^b {\partial\over\partial x^a}
           + {1\over 2} f_{AB}^C \hat{c}^A \hat{c}^B {\partial\over\partial \hat{c}^C}
           \\\hat{\Psi}\;=\;
 &
            \sum_{n\geq 1}(\Theta\langle\hat{c}\rangle(\alpha))^I_{b_1\cdots b_n} x^{b_1}\cdots x^{b_n}
            {\partial\over\partial \alpha^I}
            +
            \sum_{n\geq 2}(\Psi\langle\hat{c}\rangle(\alpha))^a_{b_1\cdots b_n} x^{b_1}\cdots x^{b_n}
            {\partial\over\partial x^a}
            \end{align}
Eq. (\ref{CommutatorInIO}) implies:
\begin{equation}
                    \hat{Q} \hat{\Psi} + {1\over 2}[\hat{\Psi},\hat{\Psi}] = 0
                    \end{equation}\label{BigMC}

\subsubsection{Reduction to ${\bf g}\subset {\bf a}$}\label{sec:VicinityOfPoint}

We want to reduce from $\bf a$ to $\bf g$, for the following reasons:

\begin{itemize}
\item The $\hat{Q}$ of Eq. (\eqref{BigMC}) is too complicated. Instead of acting
   on functions of $x$, it acts on functions on the first infinitesimal neighborhood of $\cal O$.
   The expansion in powers of of $x$ is tensored with arbitrary functions of $\alpha^I$.
\item In principle, $\bf a$ is ``implementation-dependent''; different descriptions of supergravity may have
   slightly different gauge symmetries
\item On the field theory side, we only have $\bf g$ and not $\bf a$
\end{itemize}
Therefore, we will now investigate the reduction to ${\bf g}\subset {\bf a}$. We will start by concentrating
on the vicinity of the point $p\in {\cal O}$.

Let us concentrate on the vicinity of the point $p\in \cal O$. Suppose that at the point $p$: $\alpha^I=0$.
Again, we will separate $v\langle\xi\rangle$ into linear and non-linear part.
For $\xi\in\bf g$ (\textit{i.e.} the stabilizer of $p$),  $v\langle\xi\rangle(p)=0$, therefore:

\begin{align}
  F^{-1}_{\epsilon *}v\langle\xi\rangle\;=\;
 &q\langle\xi\rangle
           + \sum_{m+n\geq 2\atop m\geq 1} \psi^a_{a_1\cdots a_m,I_1\cdots I_n}
   x^{a_1}\cdots x^{a_m}\alpha^{I_1}\cdots\alpha^{I_n}{\partial\over\partial x^a} +
  \nonumber\\ 
  & \phantom{q\langle\xi\rangle}
    + \sum_{m+n\geq 2} \theta^I_{a_1\cdots a_m,I_1\cdots I_n}
   x^{a_1}\cdots x^{a_m}\alpha^{I_1}\cdots\alpha^{I_n}{\partial\over\partial \alpha^I}
  \label{vNearP}\\ 
  \mbox{}
 &\mbox{where}\nonumber\\q\langle\xi\rangle\;=\;
 &
           \rho_{gauge}\langle\xi\rangle^I_J \alpha^J {\partial\over\partial\alpha^I} +
           \rho_{phys}\langle\xi\rangle^a_b x^b {\partial\over\partial x^a} +
           \theta\langle\xi\rangle^I_a x^a {\partial\over\partial \alpha^I}
           \end{align}
Schematically:
\begin{align}q\langle\xi\rangle\;=\;
 &[x \; \alpha]
           \left[\begin{array}{cc}
                       \rho_{phys} & \theta
                       \cr
                       0           & \rho_{gauge}
                       \end{array}
                       \right]
           \left[\begin{array}{c} \partial_x \cr \partial_{\alpha} \end{array}\right]
           \end{align}
This defines the extension of the physical states with gauge transformations:
\begin{align}\mbox{}
 &0\longrightarrow {{\bf a}\over {\bf g}} \longrightarrow T_p M \longrightarrow {\cal H}\longrightarrow 0\label{ObstacleToNI}\\\mbox{where }
 &{\cal H} = N_pM\\\mbox{}
 &{{\bf a}\over {\bf g}} = T_p{\cal O}\end{align}
The tensor $\theta\langle c\rangle^I_a$ defines a cocycle --- an element of
$C^1({\bf g}, \mbox{Hom}({\cal H}, {{\bf a} / {\bf g}}))$. Its class in
$\mbox{Ext}^1_{\bf g}\left({\cal H}, {{\bf a} / {\bf g}}\right)$ is the obstacle to finding a $\bf g$-invariant map $N_p{\cal O}\longrightarrow T_pM$.
If this obstacle is zero, then by a linear  coordinate redefinition we can put $\theta^I_a=0$, \textit{i.e.}
$q\langle\xi\rangle = \rho_{gauge}\langle\xi\rangle^I_J \alpha^J {\partial\over\partial\alpha^I} +
        \rho_{phys}\langle\xi\rangle^a_b x^b {\partial\over\partial x^a}$.
In fact, we want to remove all $\theta^I_{a_1\cdots a_m}$, not only $\theta^I_a$.
The first non-vanishing coefficient $\theta^I_{a_1\cdots a_m}$ defines a class in:
\begin{equation}\label{ObstaclesToSplit}
   \mbox{Ext}_{\bf g}^1\left(S^n{\cal H},{{\bf a} / {\bf g}}\right)
   \end{equation}
If nonzero, this is a cohomological invariant.
It is not clear to us how to interpret such
invariants on the field theory side.
It was proven in \cite{Mikhailov:2011si} that the sequence \eqref{ObstacleToNI} splits for $\cal H$ of large
enough spin. In Section \ref{sec:BetaDef} we will prove that it splits for beta-deformation, which is
the representation of the smallest nonzero spin. 
Motivated by these observations,  we will assume in the following discussion that \eqref{ObstacleToNI} always splits.
Moreover, we assume that all the obstacles in \eqref{ObstaclesToSplit} vanish. 
(Not that the whole cohomology group $\mbox{Ext}_{\bf g}^1\left(S^n{\cal H},{{\bf a} / {\bf g}}\right)$ vanishes, but the actual invariant is zero.) If true, this is a nonlinear analogue of the covariance property of the vertex
discussed in \cite{Mikhailov:2011si}. 

\paragraph     {Assuming that there are no obstacles of the type Eq. \eqref{ObstaclesToSplit}}

\begin{figure}[!htb]
     \center{\includegraphics[scale=0.3]
     {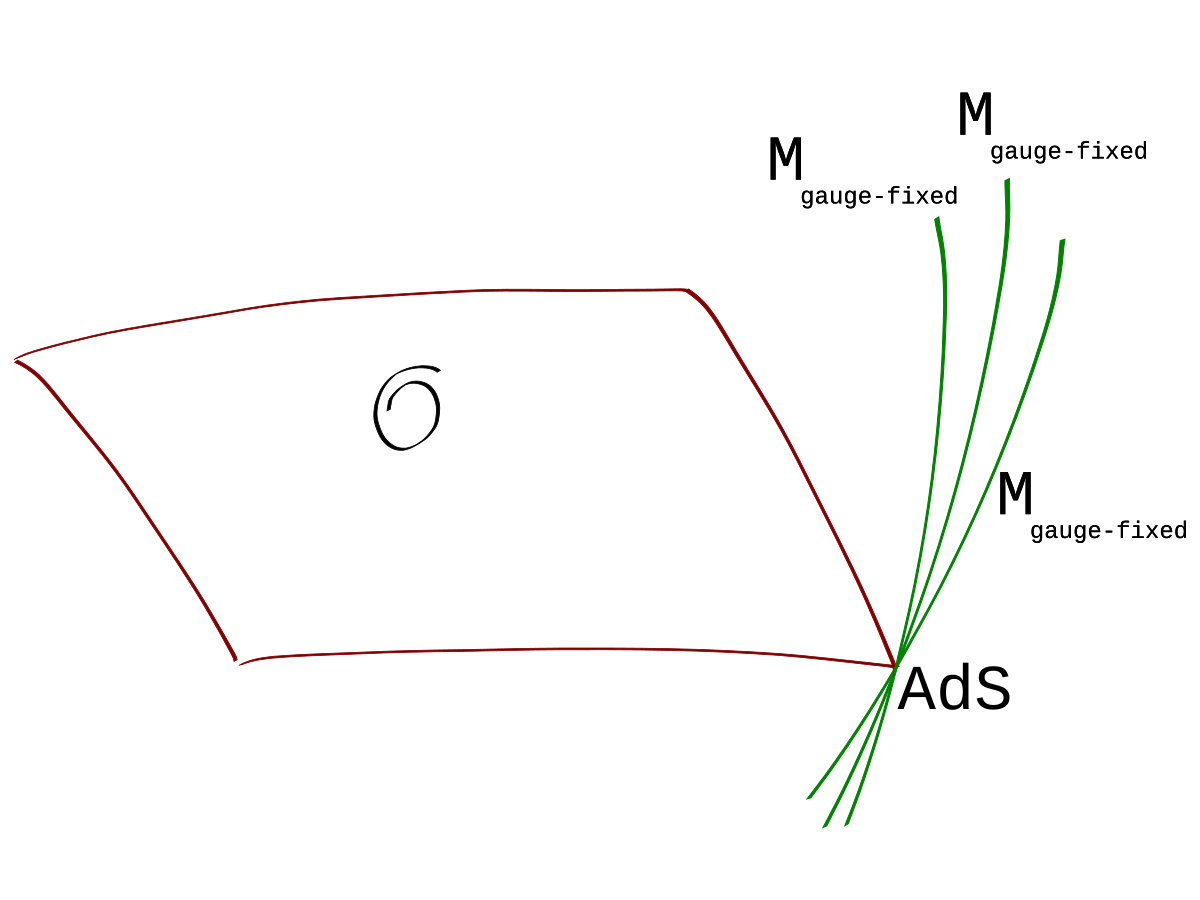}}
     \caption{\label{fig:orbit} $\cal O$ and $M_{gauge-fixed}$}
\end{figure}

If the obstacles of the type Eq. \eqref{ObstaclesToSplit} are all zero up to some value of $n$,
then we can remove all $\theta^I_{a_1\cdots a_m,I_1\cdots I_n}$ with $n=0$, so we are left with:

\begin{align}
  F^{-1}_{\epsilon *}v\langle\xi\rangle\;=\;
  &q\langle\xi\rangle
    + \sum_{m+n\geq 2\atop m\geq 1} \psi\langle\xi\rangle^a_{a_1\cdots a_m,I_1\cdots I_n}
    x^{a_1}\cdots x^{a_m}\alpha^{I_1}\cdots\alpha^{I_n}{\partial\over\partial x^a} \;+
    \nonumber\\   
  & \phantom{q\langle\xi\rangle}
    + \sum_{m+n\geq 2\atop n\geq 1} \theta\langle\xi\rangle^I_{a_1\cdots a_m,I_1\cdots I_n}
    x^{a_1}\cdots x^{a_m}\alpha^{I_1}\cdots\alpha^{I_n}{\partial\over\partial \alpha^I}
\end{align}
Geometrically this means  that we can find a $\bf g$-invariant submanifold $M_{gauge-fixed}$ which 
intersects $\cal O$ transversally at the point $p$. It is given by the equation $\alpha^I = 0$.
In this case, $v\langle\xi\rangle$ for $v\in\bf g$ defines a vector field on $M_{gauge-fixed}$,
and a solution of the Maurer-Cartan equation.
Let $c\in \Pi{\bf g}$ be the Faddeev-Popov ghost for ${\bf g}$. Restriction of $v\langle c\rangle$
to  $M_{gauge-fixed}$ is:
\begin{align}\mbox{}
 &q_{phys} + \Psi\\\mbox{where $q_{phys}\;=\;$}
 &\rho_{phys}\langle c\rangle^a_b x^b {\partial\over\partial x^a}\\\Psi \;=\;
 &\sum_{n\geq 2}\psi\langle c\rangle^a_{a_1\cdots a_n} x^{a_1}\cdots x^{a_n} {\partial\over\partial x^a}\end{align}
Let us denote:
\begin{equation}
   Q_{phys} = q_{phys} + {1\over 2}f^C_{AB}c^A c^B{\partial\over\partial c^C}
   \end{equation}
We then observe that $\Psi$ satisfies the Maurer-Cartan equation:
\begin{align}\mbox{}
 &Q_{phys}\Psi + {1\over 2}[\Psi,\Psi] = 0\label{MCPhys}\end{align}

\subsubsection{Is $\Psi$ an invariant? Deformations of $M_{gauge-fixed}$.}\label{sec:Complication}

Therefore, an action of $A$ on the vicinity of ${\cal O}\subset M$ defines some solution
$\Psi$ of the Maurer-Cartan Eq. \eqref{MCPhys}. A gauge transformation $\delta \Psi = Q_{phys}\Phi + [\Psi,\Phi]$
corresponds a change of coordinates on $M_{gauge-fixed}$.

But is it true that $\Psi$ modulo such gauge transformations is an invariant?
Generally speaking the answer is negative, for the following reason:
$M_{gauge-fixed}$ may be deformable,  different choices of $M_{gauge-fixed}$  leading to different $\Psi$,
apriori not related by a gauge transformations of the form $\delta \Psi = Q_{phys}\Phi + [\Psi,\Phi]$.

Therefore, we have to study the deformations of $M_{gauge-fixed}$. 

\paragraph     {Geometrical picture}

Deformations of a $\bf g$-invariant submanifold $M_{gauge-fixed}$
are given by $\bf g$-invariant sections of its normal bundle $N M_{gauge-fixed}$.
An exact sequence
\begin{equation}
   0 \longrightarrow TM_{gauge-fixed} \longrightarrow TM \longrightarrow NM_{gauge-fixed} \longrightarrow 0
   \end{equation}
implies the existence of a canonical map\footnote{
 $\mbox{im}\delta$ is the kernel of the natural map
$
   H^1({\bf g}, \mbox{Vect}(M_{gauge-fixed}))\longrightarrow H^1({\bf g}, \Gamma(TM|_{M_{gauge-fixed}}))
   $
}:
\begin{equation}\label{DeltaMorphism}
   \delta\;:\; H^0({\bf g}, \Gamma(NM_{gauge-fixed})) \longrightarrow H^1({\bf g}, \mbox{Vect}(M_{gauge-fixed})) 
\end{equation}
The space $H^1({\bf g}, \mbox{Vect}(M_{gauge-fixed}))$ can be naturally identified with the
tangent space to the space of solutions of the Maurer-Cartan Eq. \eqref{MCPhys} modulo gauge transformations:
\begin{equation}
   T(MC) = H^1({\bf g}, \mbox{Vect}(M_{gauge-fixed}))
   \end{equation}
Therefore, when $\delta\neq 0$, solution of MC equation depends on the choice of $M_{gauge-fixed}$.
This means that we have, generally speaking, a {\em family} of invariants.

\paragraph     {In coordinates}

Remember that $M_{gauge-fixed}$ is given by the equations:
\begin{equation}
   \alpha^I=0
   \end{equation}
Infinitesimal deformations of  $M_{gauge-fixed}$ can be obtained as fluxes by vector fields of the form
\begin{equation}
   Y^I_{a_1\cdots a_n}x^{a_1}\cdots x^{a_n}{\partial\over\partial\alpha^I} + \ldots 
   \end{equation}\label{DeformingVectorFieldLeadingTerm}
where $Y^I_{a_1\cdots a_n}x^{a_1}\cdots x^{a_n}{\partial\over\partial\alpha^I}$ commute with
$\rho_{gauge}\langle\xi\rangle^I_J \alpha^J {\partial\over\partial\alpha^I} +
        \rho_{phys}\langle\xi\rangle^a_b x^b {\partial\over\partial x^a}$ .
In other words, the tensor field $Y_{a_1\cdots a_n}^I$ defines an element of $\mbox{Hom}_{\bf g}(S^n{\cal H}, {\bf a}/{\bf g})$.
The flux of such vector fields preserves the condition that  $\theta^I_{a_1\cdots a_k} = 0$ for $k\leq n$.
To keep  $\theta^I_{a_1\cdots a_k} = 0$ for $k>n$, we need to add to (\eqref{DeformingVectorFieldLeadingTerm})
some terms of higher order in $x$ (the $\ldots$ in Eq. (\eqref{DeformingVectorFieldLeadingTerm}));
the existence of such terms depends on the vanishing of  $\mbox{Ext}^1(S^n{\cal H}, {\bf a}/{\bf g})$.

For example, consider a vector field $Y^{(0)}$:
\begin{equation}
   Y^{(0)} = y^I_a x^a {\partial\over\partial\alpha^I}
   \end{equation}
where $y^I_a$ is a $\bf g$-invariant tensor in $\mbox{Hom}({\cal H}, {\bf a}/{\bf g})$.
The commutator $\left[ v\langle\xi\rangle\,,\,Y^{(0)}\right]$ contains terms:
\begin{equation}
   u\langle\xi\rangle = \left(
         \psi\langle\xi\rangle^a_{a_1 a_2}y^{I}_a - \theta\langle\xi\rangle^I_{a_1,I_1} y^{I_1}_{a_2}
         \right) x^{a_1}x^{a_2} {\partial\over\partial \alpha^I}
   \end{equation}
They automatically satisfy:
\begin{equation}
   [q\langle\xi\rangle,u\langle\eta\rangle] - (\xi\leftrightarrow\eta) - u\langle[\xi,\eta]\rangle
   = 0
   \end{equation}
and under the assumption of vanishing $\mbox{Ext}^1(S^2{\cal H}, {\bf a}/{\bf g})$ exists $Y^{(1)}$:
\begin{align}Y^{(1)}\;=\;
 &y_{a_1a_2}^{I} x^{a_1}x^{a_2}{\partial\over\partial \alpha^I}\\u\langle\xi\rangle \;=\;
 &[q\langle\xi\rangle, Y^{(1)}]\end{align}
Therefore we can construct, order by order in $x$-expansion, a vector field:
\begin{equation}
   Y = \left(y^{I}_a x^a + y^{I}_{ab} x^a x^b + \ldots\right)
   {\partial\over\partial\alpha^I}
   \end{equation}
defining a $\bf g$-invariant section of the normal bundle of $M_{gauge-fixed}$, and
therefore a deformation of $M_{gauge-fixed}$ as a $\bf g$-invariant submanifold.
The corresponding deformation of $\Psi$ is:
\begin{align}\delta \Psi = [v\langle c\rangle, Y]|_{\alpha = 0}\;=\;
 &\sum_{m\geq 1}\sum_{n\geq 1}
                      \psi\langle c\rangle^a_{a_1\cdots a_m,I}\;y^{I}_{a_{m+1}\cdots a_{m+n}}\;
                      x^{a_1}\cdots x^{a_{m+n}}{\partial\over\partial x^a}
\end{align}
We do not see any apriori reason why this  $\delta\Psi$ could be absorbed by a gauge
transformation, {\it i.e.} why would exist $\Phi$ such that $\delta \Psi = Q_{phys}\Phi + [\Psi,\Phi]$.

\subsubsection{Conclusion}\label{sec:AdSAbstractConclusion}

We have studied the problem of classifying the normal forms of the action of a Lie supergroup $A$
in the vicinity of an orbit $\cal O$ with nontrivial stabilizer.
As invariants of the action, we found families of equivalence classes
of solutions of MC equations modulo gauge transformations.
These families are parameterized by $\bf g$-invariant submanifolds $M_{gauge-fixed}$.

Remember that $\bf g$ is a finite-dimensional
Lie superalgebra, while $\bf a$ is infinite-dimensional, and therefore $\cal O$ is infinite-dimensional.
One would think that
the deformations of $M_{gauge-fixed}$ ``along $\cal O$'' will be ``as complicated as $\cal O$''.
That would be ugly. But in fact, the space of deformations
of $M_{gauge-fixed}$ is ``no more complicated than $\cal H$''. For examle, when $\cal H$ is
finite-dimensional, the space of deformations is also finite-dimensional at each order in $x$.
Roughly speaking, at the order $n$, it is  $\mbox{Hom}_{\bf g}(S^n{\cal H}, {\bf a}/{\bf g})$.
Even though $\mbox{Hom}_{\bf g}(S^n{\cal H}, {\bf a}/{\bf g})$
may be non-zero, it is certainly nicer than ${\bf a}/{\bf g}$.

In other words, the deformations of $M_{gauge-fixed}$ do  not involve the dependence on all $\alpha^I$,
but only on  finite-dimensional subspaces, the images of intertwining operators
$S^n{\cal H} \longrightarrow {\bf a}/{\bf g}$.

\subsection{Pure spinor formalism}\label{PureSpinor}

Here we will briefly outline the pure spinor implementation of supergravity in the vicinity
of AdS. 
We use the notations of \cite{Mikhailov:2011si}.
Let ${\bf g}_0\subset {\bf g}$ be the subalgebra preserving a point in $AdS_5\times S^5$. 
In the pure spinor formalism, the space of vertex operators (cochains) of ghost number $n$ transforms
in an induced representation of $\bf g$:
\begin{equation}
   C_{ps}^n = \mbox{Coind}_{{\bf g}_0}^{\bf g} {\cal P}^n
\end{equation}
where ${\cal P}^n$ is the space of homogeneous polynomials of the order $n$ on the pure spinor variable.
The space of linearized SUGRA solutions corresponds to the cohomology at ghost number $n=2$:
\begin{align}T_p M \;=\;
 &Z^2_{ps} = \mbox{ker}\;:\; C_{ps}^2 \longrightarrow C_{ps}^3\\T_p {\cal O} \;=\;
 &B^2_{ps} = \mbox{im}\;:\; C_{ps}^1 \longrightarrow C_{ps}^2\\{\cal H}\;=\;
 &H_{ps}^2 = Z_{ps}^2/B_{ps}^2\end{align}
This defines an extension of $\cal H$ with $T_p {\cal O}$:
\begin{align}\mbox{}
  &0 \longrightarrow B^2_{ps} \longrightarrow Z^2_{ps} \longrightarrow {\cal H} \longrightarrow 0
    \label{ExtensionOfH}
\end{align}
The structure of extension is described by a cocycle
\begin{equation}
   \alpha\in Z^1({\bf g}, \mbox{Hom}({\cal H}, B^2_{ps}))
\end{equation}
The extension is nontrivial iff $\alpha$ defines a nontrivial class in $\mbox{Ext}^1_{\bf g}({\cal H}, B^2_{ps})$.
This class is an obstacle to the existence of a covariant vertex.

We have seen (Section \ref{sec:VicinityOfPoint}) that, more generally, $\mbox{Ext}^1_{\bf g}(S^n{\cal H}, B^2_{ps})$
are obstacles to finding a $\bf g$-invariant gauge for nonlinear solutions.
We will now show that $\mbox{Ext}^1_{\bf g}(S^n{\cal H}, B^2_{ps})$, although nonzero, is in some sense small.

Let us denote:
\begin{equation}
   V = S^n{\cal H}
\end{equation}
 Notice that $B^2_{ps}$ fits in a short exact sequence or representations,
with the corresponding long exact sequence of cohomologies:
\begin{align}\mbox{}
 &0 \longrightarrow Z^1_{ps} \longrightarrow C^1_{ps} \longrightarrow B^2_{ps} \longrightarrow 0\\\mbox{}
 & \mbox{Ext}^1_{\bf g}(V,C^1_{ps})
   \longrightarrow \mbox{Ext}^1_{\bf g}(V,B^2_{ps})
   \longrightarrow \mbox{Ext}^2_{\bf g}(V,Z^1_{ps})
   \longrightarrow \mbox{Ext}^2_{\bf g}(V,C^1_{ps})
\end{align}
Shapiro's lemma implies that $\mbox{Ext}^1_{\bf g}(V,C^1_{ps})=  \mbox{Ext}^2_{\bf g}(V,C^1_{ps}) = 0$.
For example, for $\mbox{Ext}^1_{\bf g}(V,C^1_{ps})$ we have:
\begin{equation}
   \mbox{Ext}^1_{\bf g}(V,C^1_{ps}) =   \mbox{Ext}^1_{{\bf g}_0}(V|_{{\bf g}_0}, {\cal P}^1) =  0
\end{equation}
because $V|_{{\bf g}_0}$ is semisimple as a representation of ${\bf g}_0$, and $H^1({\bf g}_0)=0$.
Therefore:
\begin{equation}
   \mbox{Ext}^1_{\bf g}(V, B^2_{ps}) = \mbox{Ext}^2_{\bf g}(V,Z^1_{ps})
\end{equation}
Notice that $Z^1$ is in the following exact sequences
(the pure spinor cohomology at ghost number one, $H^1_{ps}$, corresponds to the global symmetries $\bf g$.):
\begin{align}\mbox{}
 &0\longrightarrow B^1_{ps} \longrightarrow Z^1_{ps} \longrightarrow [H^1_{ps} = {\bf g}]\longrightarrow 0\\\mbox{}
 &\mbox{Ext}^1_{\bf g}(V,{\bf g})
      \longrightarrow \mbox{Ext}^2_{\bf g}(V,B^1_{ps})
      \longrightarrow \mbox{Ext}^2_{\bf g}(V,Z^1_{ps})
      \longrightarrow \mbox{Ext}^2_{\bf g}(V,{\bf g})
      \longrightarrow \mbox{Ext}^3_{\bf g}(V,B^1_{ps})
\end{align}
BRST exact vertices of ghost number one, {\it i.e.}  $B^1_{ps}$, fit in the following exact sequences:
\begin{align}\mbox{}
 &0\longrightarrow {\bf C} \longrightarrow C^0_{ps} \stackrel{Q}{\longrightarrow} B^1_{ps}\longrightarrow 0\\\mbox{}
 & \mbox{Ext}^2_{\bf g}(V,C^0_{ps})
   \longrightarrow \mbox{Ext}^2_{\bf g}(V,B^1_{ps})
   \longrightarrow \mbox{Ext}^3_{\bf g}(V,{\bf C})
   \longrightarrow \mbox{Ext}^3_{\bf g}(V,C^0_{ps})
\end{align}
These observations together imply that $\mbox{Ext}^1_{\bf g}(V, B^2)$ fits into a short exact sequence of linear
spaces:
\begin{equation}
   0 \longrightarrow
   {\mbox{ker}\left[\mbox{Ext}^3_{\bf g}(V, {\bf C})\longrightarrow \mbox{Ext}_{{\bf g}_0}^3(V, {\bf C})\right]
     \over \mbox{im}\left[\mbox{Ext}^1_{\bf g}(V,{\bf g})\rightarrow\mbox{Ext}^2_{\bf g}(V, B^1)\right]} \longrightarrow \mbox{Ext}^1_{\bf g}(V, B^2) \longrightarrow
   \mbox{Ext}^2_{\bf g}(V, {\bf g}) \longrightarrow 0
\end{equation}
This means that $\mbox{Ext}^1_{\bf g}(V, B^2)$ is a direct sum:
\begin{align}
  \mbox{Ext}^1_{\bf g}(V, B^2)\;=\;
  & E \oplus \mbox{Ext}^2_{\bf g}(V, {\bf g})
  \\
  \mbox{\tt where } E\;=\;
  & {\mbox{ker}\left[\mbox{Ext}^3_{\bf g}(V, {\bf C})\longrightarrow \mbox{Ext}_{{\bf g}_0}^3(V, {\bf C})\right]
    \over \mbox{im}\left[\mbox{Ext}^1_{\bf g}(V,{\bf g})\rightarrow\mbox{Ext}^2_{\bf g}(V, B^1)\right]}
\end{align}
Notice that $E$ is a factorspace of a subspace of $\mbox{Ext}^3_{\bf g}(V, {\bf C})$.
In this sense, $\mbox{Ext}^1_{\bf g}(S^n{\cal H}, B^2)$ is ``lesser than'':
\begin{equation}\label{CohomologiesOfGaugePart}
\mbox{Ext}^3_{\bf g}(S^n{\cal H}, {\bf C}) \oplus \mbox{Ext}^2_{\bf g}(S^n{\cal H}, {\bf g})
\end{equation}
This is the ``upper limit'' on the cohomology group containing the obstacle to
the existence of $M_{gauge-fixed}$ at the order $n$ in $x$-expansion.

\paragraph     {When $n=1$}
In particular $n=1$ corresponds to the obstacle to the existence of a covariant vertex,
{\it i.e.} to the splitting of the short exact sequence of Eq. (\ref{ExtensionOfH}).
In Section \ref{sec:BetaDef} we will show that  the actual obstacle is zero in the case of
linearized beta-deformation (which is the representation with smallest spin).
At the same time, results of \cite{Mikhailov:2011si} imply that the obstacle is zero for
linearized solutions with large enough spin. The conjectured existence of covariant vertices
is the interpolation between these two cases.

\subsection{If $M_{gauge-fixed}$ does not exist}

We do not have a proof that the obstacle in $\mbox{Ext}_{\bf g}^1\left(S^n{\cal H},{{\bf a} / {\bf g}}\right)$
defined in Section \ref{sec:VicinityOfPoint} is actually zero.
If it is not zero, then we cannot restrict to $M_{gauge-fixed}$ (because  $M_{gauge-fixed}$ does not exist).
Then we must study the full expansion of $F_{\epsilon *}^{-1}v$ of Eq. (\ref{vNearP}), in powers of {\em both}
$x^a$ {\em and} $\alpha^I$.

However, we do not need to take into account all $\alpha^I$. We only need those $\alpha^I$ which
represent the obstacle in $\mbox{Ext}_{\bf g}^1\left(S^n{\cal H},{{\bf a} / {\bf g}}\right)$. Therefore,
the complication is not actually as bad as it could have been. The main point is that,
even though ${\bf a}/{\bf g}$ seems  an ugly infinite-dimensional space, its cohomologies
are reduced to expressions like (\ref{CohomologiesOfGaugePart}) 
by the ``magic'' of the Shapiro's lemma.

\vspace{10pt}

\paragraph     {In the rest of this Section} we will accept as a working hypothesis that $M_{gauge-fixed}$ exists.
Also, we will leave open the question of non-uniqueness of $M_{gauge-fixed}$.

\subsection{Normalizable SUGRA solutions}\label{sec:NormalizableSolutions}
``Normalizable'' means decreasing sufficiently rapidly near the boundary.

\vspace{10pt}
\noindent
All {\em linearized}  normalizable SUGRA solutions are periodic in the global time $t$.
They approximate some complete (nonlinear) solutions. The nonlinear solutions are {\em not} periodic.
But, since linearized solutions {\em are} periodic, we can define the monodromy transformation
$m$ as in Section \ref{sec:MonodromyTransformation}.

The space of normalizable ({\it i.e.} rapidly decreasing at the boundary) solutions
has a symplectic form. This is true at the linearized level as well as for the
non-linear solutions. Then we can choose a map $T_0 M\rightarrow M$ so that it preserves
the symplectic structure\footnote{This can be done, for example, in the following way.
Pick some timelike surface. Then, to every linearized SUGRA solution associate
a nonlinear solution for which the values of SUGRA fields and their time derivatives
at that timelike surface are same as for the linearized solution}.
Therefore, we can now identify $\Psi$ with the corresponding Hamiltonian, which we denote $H_{\Psi}$,
or just $H$. The MC equation (\ref{MaurerCartan}) becomes:
\begin{equation}
QH + {1\over 2} \{H,H\} = 0
\end{equation}
Remember that $H$ is of cubic and higher order in the coordinates on $T_0M$.

Given the monodromy matrix of Eq. (\ref{MonodromyTransformation}) we can (in perturbation theory)
define a vector field $\xi$ such whose flux generates it:
\begin{equation}
e^{\xi} = m
\end{equation}
It has some Hamiltonian $H_{\xi}$ which is of cubic and higher order
in the coordinates on $T_0M$. In some sense, the quantization of $H_{\xi}$ should give the
spectrum of anomalous dimensions. This program is complicated, though, by a non-straigthforward action
of the symmetry, see Section \ref{sec:Symmetries}.

\subsection{Non-normalizable SUGRA solutions}
The non-normalizable SUGRA solutions correspond to the deformations of the boundary theory.
Consider the following element of $PSU(2,2|4)$ --- the symmetry group of $AdS_5\times S^5$:
\begin{equation}
S=\mbox{diag}(i,i,i,i,-i,-i,-i,-i)\in PSU(2,2|4)
\end{equation}
Suppose that our deformation is invariant under $(-1)^FS$:
\begin{equation}\label{SInv}
U = (-1)^FSU
\end{equation}
Then, the corresponding {\em linearized} solution is periodic in the global time of AdS.
\begin{figure}[!htb]
     \center{\includegraphics[width=\textwidth]
     {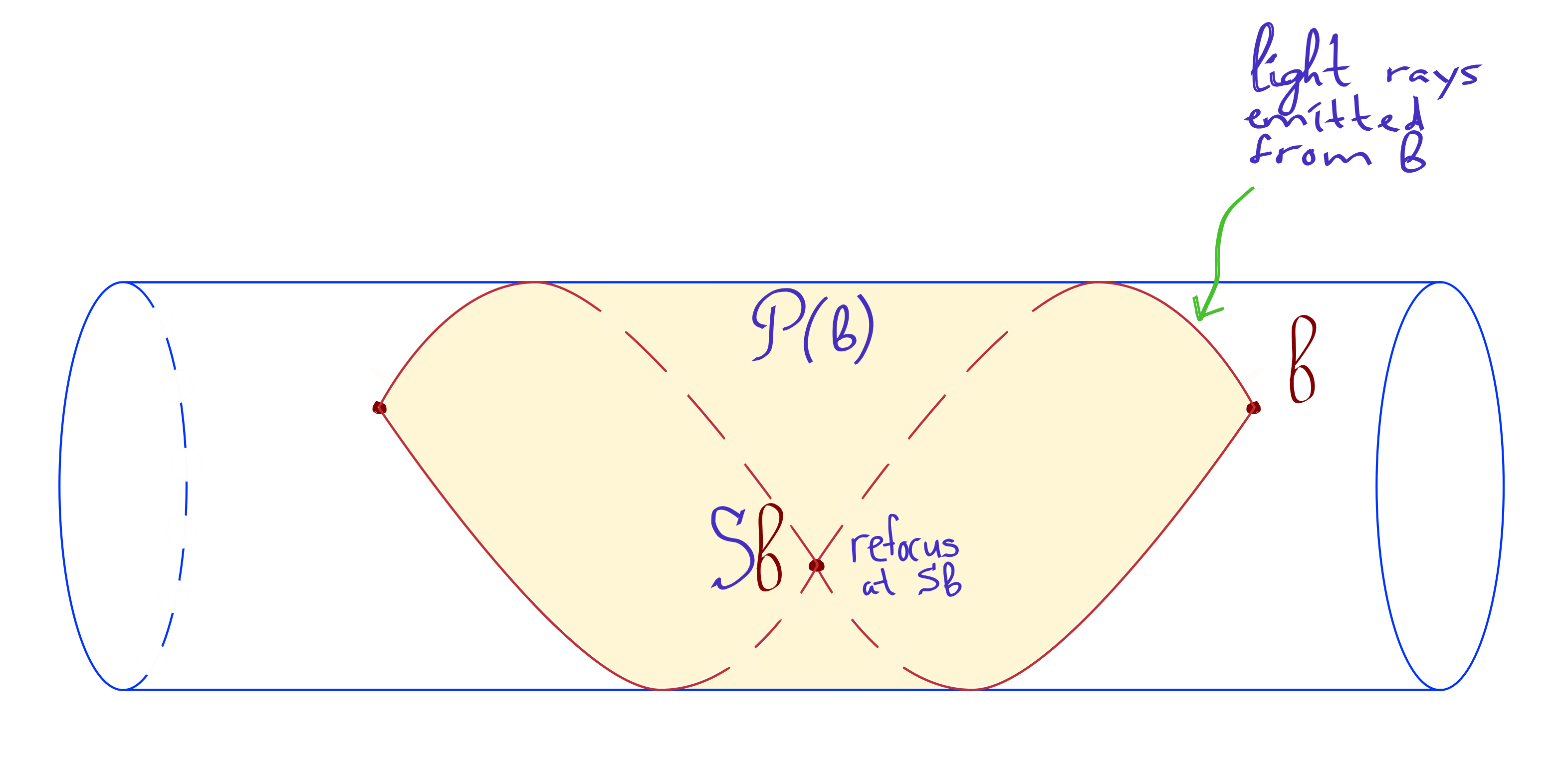}}
     \caption{\label{fig:defS} Transformation $S$, as it acts on the boundary of AdS}
\end{figure}
Indeed, let us consider the retarded wave generated by an insertion of some operator $\cal O$
at the point $b$ on the boudary. It is the same as retarded boundary-to-bulk propagator.
It is a generalized function with support on the future light cone of $b$.
Let $l\in {\bf R}^{2+4}$ be a light-like vector encoding the point $b$ on the boundary.
Then the boundary-to-bulk propagator, in sufficiently small neighborhood of $b$ is, schematically:
\begin{equation}\label{BoundaryToBulkPropagator}
{\delta(v\cdot l)\over (v\cdot l)^{\Delta -1}}
\end{equation}
where $v\in {\bf R}^{2+4}$ with $(v,v)=1$ corresponds to a point inside AdS;
$\Delta$ is the conformal dimension of $\cal O$. The future light cone gets re-focused
at $Sb$. The free solution (\ref{BoundaryToBulkPropagator}) gets then reflected from
the boundary at the point $Sb$, and when reflected changes sign. Therefore, in order to
cancel the reflection, we have to put the same operator at the point $SU$. 

\vspace{10pt}
\noindent
However, the corresponding nonlinear solution may or may not be periodic.
If it is not periodic, then the deviation from
periodicity is characterized by the monodromy of Eq. (\ref{MonodromyTransformation}).
In any case, the solution of the Maurer-Cartan equation is more fundamental 
than the monodromy transformation.

\subsection{Simplest non-periodic linearized solution}
(This subsection is a side remark.)

\noindent
As we mentioned in Section \ref{sec:NormalizableSolutions}, all {\em normalizable} linearized
solutions are periodic in global time $t$.

But of course, this is not true for non-normalizable solutions. (Indeed, nothing prevents
us from considering non-periodic boundry conditions at the boundary of AdS. There exist
corresponding solutions, which are not periodic.) As a simplest example, consider the dilaton
linearly dependent on $t$:
\begin{equation}\label{LinearDilaton}
   \phi = \alpha  t
   = {\alpha\over 2i} \log {Z\over\overline{Z}}
   \quad , \quad \alpha=\mbox{const}
\end{equation}
This is a solution of SUGRA only at the linearized level. Indeed, the energy $(\dot{\phi})^2$ is
nonzero, and it will deform the metric. It would be interesting to see if it approximates
some solution of nonlinear equations with the following property: the action of $\partial\over\partial t$ on it
is the shift of dilaton.

If we act on $\phi$ of Eq. (\ref{LinearDilaton}) by generators of $\bf g$,
we get an infinite-dimensional representation.
This infinite-dimensional representation contains a 1-dimensional invariant subspace, because
the action of $\partial\over\partial t$ gives constant.
(But the action of $K_i$ and $\bar{K}_i$ of Eqs. (\ref{GenK}), (\ref{GenBarK}) results
in expressions like $X_i\over Z$, {\it etc}, an infinite-dimensional space.)

\section{Beta-deformation and its generalizations}\label{sec:BetaDeformation}
We will now consider the case of  beta-deformation. See \cite{Leigh:1995ep,Milian:2016xuy}
for the description on the field theory side,
and \cite{Bedoya:2010qz,Benitez:2018xnh}
for the AdS description\footnote{Particular ``subsectors'' of AdS beta-deformations
  were described earlier in \cite{Fayyazuddin:2002vh,Aharony:2002hx,Lunin:2005jy,Chen:2006bh}.
  While most of work has been on special cases associated to Yang-Baxter equations,
  the authors of \cite{Aharony:2002hx} studied more generic
  values of the parameter, which have nonzero beta-funcion. In this work we consider
  most general values of the deformation parameter.
}.
It does satisfy Eq. (\ref{SInv}).
Linearized beta-deformations transform in the following representation:
\begin{equation}\label{DefCalH}
{\cal H} = {({\bf g}\wedge {\bf g})_0\over {\bf g}}
\end{equation}
where the subindex $0$ means zero internal commutator in the central extended $\hat{\bf g}$.
This means
$x\wedge y\in ({\bf g}\wedge {\bf g})_0$ has $[x,y]=0$ where the commutator is taken in $\hat{\bf g}$
({\it i.e.} the unit matrix is not discarded).

It was shown in \cite{Aharony:2002hx} that the renormalization
of beta-deformation is again a beta-deformation, and the anomalous
dimension is an expression cubic in the beta-deformation parameter. 

We will conjecture that 
the expansion of $\Psi$ starts with quadratic terms, and not with cubic terms.
This explains why the obstacle found in \cite{Aharony:2002hx}    is actually
quadratic rather than cubic.

But first,
in order to make contact with Sections \ref{sec:GeometricaAbstractionAdS}, \ref{sec:NormalFormA}, \ref{PureSpinor}
we will discuss the description of the beta-deformation in the pure spinor formalism.

\subsection{Description of beta-deformation in pure spinor formalism}
\label{sec:BetaDef}

Vertex operators of physical states live in $C^2_{ps}$ --- cochains of ghost number two.
The vertex corresponding to beta-deformation, as constructed in \cite{Mikhailov:2011si,Bedoya:2010qz} is
actually not covariant. It transforms in $({\bf g}\wedge {\bf g})_0$ instead of Eq. (\ref{DefCalH}).
Some components of the vertex, transforming in $\bf g$, are BRST exact:
\begin{equation}
   \begin{array}{rcccccl}
   &  0          & & 0          & & &
   \cr
   &  \downarrow & & \downarrow & & &
   \cr
   0 \longrightarrow & {\bf g} & \stackrel{i}{\longrightarrow} & ({\bf g}\wedge {\bf g})_0 & \longrightarrow &
   {({\bf g}\wedge {\bf g})_0\over {\bf g}} & \longrightarrow 0
   \cr
   & f\downarrow & &  j\downarrow  &  & ||  &
   \cr
   & C_{ps}^1 & \stackrel{Q_{ps}}{\longrightarrow} & C_{ps}^2 & & {\cal H} &
   \end{array}
   \end{equation}
This defines a nontrivial extension, which can be characterized by a cocycle:
\begin{equation}
   \alpha \;:\; {({\bf g}\wedge {\bf g})_0\over {\bf g}} \longrightarrow {\bf g}
   \end{equation}
defining a nonzero class in $\mbox{Ext}^1\left({({\bf g}\wedge {\bf g})_0\over {\bf g}}, {\bf g}\right)$.
The existence of the commutative square formed by $i,j,Q_{ps},f$ is nontrivial.
Then nontriviality is in the fact that $f$ commutes with the action of $\bf g$.
Generally speaking, the variation of $f$ under the action of $\bf g$ could be non-zero, it just has to take values
in $Z^1_{ps}$. But there is a $\bf g$-invariant $f$:
\begin{equation}
   f(\xi) = \mbox{STr}(g\xi g^{-1} (\lambda_L + \lambda_R))
   \end{equation}
The composition $f\circ\alpha$ defines an element $H^1({\bf g}, \mbox{Hom}({\cal H}, C_{ps}^1))$, but
this group is zero by Shapiro's lemma:
\begin{equation}
   H^1({\bf g}, \mbox{Hom}({\cal H}, C^1_{ps})) = H^1({\bf g}_0 , \mbox{Hom}({\cal H}, {\cal P}^1)) \;=\;0
   \end{equation}
(Here ${\cal P}^1$ is the space of linear functions of pure spinors.)
Therefore exists $\beta\in C^0({\bf g}, \mbox{Hom}({\cal H}, C^1_{ps}))$ such that:
\begin{equation}
   f\circ \alpha = Q_{\rm Lie}\beta
   \end{equation}
This means that the BRST-equivalent vertex:
\begin{equation}
   V' = V - Q_{ps} \beta
\end{equation}
is $\bf g$-covariant. This is {\em not} the vertex found in \cite{Mikhailov:2011si,Bedoya:2010qz}.
It is probably a linear combination of the vertex of     \cite{Mikhailov:2011si,Bedoya:2010qz}
and the one found in \cite{Flores:2019dwr}.

\subsection{Restriction of $\Psi$ to even subalgebra}
Let us start by forgetting about fermionic symmetries. In other words, consider the
restriction of $\Psi$ on the even subalgebra  ${\bf g}_{\rm ev}\subset {\bf g}$.
Explicit computations of \cite{Aharony:2002hx} suggest\footnote{although the computations was
  only done for the simplest deformation, the one constant in AdS} that $\Psi$ starts
with cubic terms, {\it i.e.} with $v_2$ (see Eq. (\ref{ExpansionOfVectorField})) rather than $v_1$.
The $v_2$ is certainly nonzero, and cannot be removed by the gauge transformations of Section \ref{sec:GaugeTransformations}.
This leads to an apparent contradiction. Indeed, $v_2$ being non-removable means that it
represents a nontrivial class in:
\begin{equation}
   H^1\left({\bf g}_{\rm ev}, \mbox{Hom}({\cal H}^{\otimes 3} , {\cal H})\right)
\end{equation}
But this cohomology group is zero, because $\mbox{Hom}\left({\cal H}^{\otimes 3} , {\cal H})\right)$
is finite-dimensional. The $H^1$ of ${\bf g}_{\rm ev}$ with coefficient in a
finite-dimensional representation is zero --- see \cite{FeiginFuchs}.
What actually happens is:
\begin{equation}\label{BosonicCocycle}
   [v_2] = \phi_{\rm bos} \in H^1\left({\bf g}_{\rm ev}, \mbox{Hom}({\cal H}^{\otimes 3} , \widehat{\cal H})\right)
\end{equation}
where $\widehat{\cal H}$ is some infinite-dimensional extension of $\cal H$.
We will now explain this.

\subsection{Infinite-dimensional extension of finite-dimensional $\cal H$}\label{sec:InfiniteDimensionalExtension}

The perturbative nonlinear solution involves terms proportional to
$\log (Z\overline{Z})$ in the notations of Appendix \ref{sec:AdSNotations}.
These terms, under the action of $K_i$ and $\overline{K}_i$ (see Eqs. (\ref{GenK}), (\ref{GenBarK})),
generate an infinite-dimensional representation. We will first explain the origin of $\log (Z\overline{Z})$-terms,
and then the structure of the infinite-dimensional extension of $\cal H$.

\paragraph     {Log terms}
We will now explain the origin of  at the third order in the deformation parameter.
Following \cite{Aharony:2002hx}, let us consider those linearized beta-deformations which
only involve the RR fields and the NSNS $B$-field in the direction of $S^5$. At the linear level, these
solutions do not deform $AdS_5$ at all\footnote{The existence of such deformations may appear
  contradictory, because $S^5$ is a compact manifold.
  Indeed, the NSNS $B$-field is a two-form. Nonzero harmonic two-form cannot exist on a compact manifold.
  But in fact, because of the undeformed $AdS_5\times S^5$ has the RR five-form turned on, the linearized
  equations actually mix RR with NSNS fields, leading effectively to massive equations. 
}.
At the cubic order, the interacting term has three linearized solutions combine in a term proportional,
again, to the beta-deformation of $S^5$. This means that we have to solve the equation in $AdS_5$:
\begin{equation}
   {\bf L}_{\rm A} \phi = 1
\end{equation}
The solution is  $\log (Z\overline{Z})$, see Eq. (\ref{LogZBarZ}).
At higher orders of $\epsilon$-expansion, more complicated function appear, see Appendix \ref{sec:BasicFunctions}.
All non-rational dependence of $Z,\overline{Z},\vec{X}$ is through $|Z|^2=Z\overline{Z}$. Denominators
are powers of $Z$ and $\overline{Z}$, while $\vec{X}$ only enter polynomially.

\paragraph     {Structure of $\widehat{\cal H}$}
(Notations of Appendix \ref{sec:AdSNotations}.)
Consider, for example, a massless scalar field, whose $S^5$-dependence corresponds
to a harmonic polynomial $Y({\bf N})$ of degree $\Delta_S$. There are solutions of the form:
\begin{equation}\label{PhiTimesY}
   \phi(Z,\overline{Z},\vec{X}) Y({\bf N})
\end{equation}
where $\phi$ is a harmonic polynomial of degree $\Delta_A = \Delta_S$.
Such solutions generate a finite-dimensional representation $V$ of $\bf g$.

Now let us allow $\phi$ to have denominators, either ${1\over Z^m}$ or $1\over\overline{Z}^m$,
keeping the same overall homogeneity degree  $\Delta_A = \Delta_S$.
(It is important that $Z$ is never zero, in fact $|Z|^2 > 1$.)
Then, the solutions (still given by Eq. (\ref{PhiTimesY})) generate an infinite-dimensional
representation $\widehat{V}$ of $\bf g$. It contains a finite-dimensional subspace
corresponding to polynomial $\phi$. Therefore, $\widehat{V}$ is an extension of $V$.
This extension is non-split\footnote{There are actually two representations, one
  allowing $1\over Z^m$ and another allowing $1\over\overline{Z}^m$. But we want to
  keep the real structure, so we must combine them.},
because there is no invariant subspace complementary to $V\subset \widehat{V}$.

We explained the construction of extension for the simplest case of the massless
scalar field. The construction for other finite-dimensional representations is the same.
Finite-dimensional $V$ involves scalar fields, tensor fields, and fermionic spinor fields,
and they all can be non-constant spherical harmonics on $S^5$.
Importantly, for a finite-dimensional $V$, all these fields are polynomials in $Z,\overline{Z},\vec{X}$.
To extend $V$ to $\widehat{V}$,  we just allow negative power of $Z$ or negative powers of $\overline{Z}$.

We would like to stress that all these extensions only contain rational functions of $Z,\overline{Z},\vec{X}$,
with denominators powers of $Z$ and $\overline{Z}$. The perturbative solution contains
non-rational terms, {\it e.g.}  $\log(Z\overline{Z})$. But  $\Psi$ only has rational functions,
all logs got differentiated out.
In this sense $\Psi$ is simpler than $\phi$ ({\it cp.} Section \ref{sec:PsiIsSimple}).

\paragraph     {Example of cocycle}
A nontrivial cocycle $\psi$  in $H^1({\bf so}(2,4),\widehat{\bf C})$ can be defined by the following formulas
(notations of Eqs. (\ref{GenK}), (\ref{GenBarK})):
\begin{align}
  & \psi\left({\partial\over\partial t}\right) \;=\; i
  \\
  & \psi(K_i) = 2{X_i\over Z}
  \\
  & \psi(\overline{K}_i) = 0
  \\
  & \psi(\mbox{\tt\small rotations of $S^3$}) = 0
\end{align}
(In other words, $\psi(x)$ is the variation of $\log Z$ under $x$.)
Here $\widehat{\bf C}$ is the infinite-dimensional extension of the trivial
representations, generated by massless scalar fields allowing denominators powers of $Z$.
Composing this $\psi$ with some intertwining operators $V^{\otimes n}\rightarrow {\bf C}$
we may get nontrivial cocycles in $H^1({\bf so}(2,4),\mbox{Hom}(V^{\otimes n},\widehat{\bf C}))$.

\subsection{Renormgroup flow generates infinite-dimensional representations}

We must stress that $\Psi$ takes values in $\widehat{\cal H}$, and not in $\cal H$.
Of course, $\cal H$ is contained in $\widehat{\cal H}$, as a subrepresentation.
But there is no invariant projector from $\widehat{\cal H}$ to $\cal H$.
We can say that the renormgroup flow of the beta-deformation results in
the infinite-dimensional extension of the representation of beta-deformation.

\vspace{10pt}
\noindent
The same is true for other finite-dimensional deformations. A renormgroup
flow of a finite-dimensional deformation generates infinite-dimensional extensions of
(possibly other) finite-dimensional representations.

\vspace{10pt}
\noindent
This, of course, implies that we should also extend our space of linearized solutions.
We should start with $\widehat{V}$ rather than just $V$. Otherwise, in the
notations of Section \ref{sec:ActionOfSymmetry}, $v$ will lead out\footnote{The rule of the game
  is to always include sufficiently large class of linearized solutions, so that the action
  of $\bf g$ on the lifted solutions does not lead out of the space of lifted solutions}
of ({\it i.e.} not tangent to)
the image of $F$.  Then, the coefficient of $\epsilon^m$ in the expansion of $\Psi$ in powers of $\epsilon$
lives in $\mbox{Hom}(\widehat{V}^{\otimes m}, \widehat{V})$.

\subsection{Nonlinear beta-deformations have trivial monodromy}\label{sec:TrivialMonodromy}
We have an intuitive argument, that the monodromy always takes values in unitary representations.
Indeed, non-normalizable excitations can be thought of as waves bouncing back and forth from
the boundary of AdS. They can be all damped by emitting appropriate excitations from the
boundary, {\it i.e.} by adjusting the boundary conditions. Only normalizable modes remain.
(See Section \ref{sec:NormalizableAndNonNormalizableMonodromy}.)

\commentstarts{\small
  For example, suppose that we need to solve the equation $\square f = Z \cos\theta$ where
  $\theta$ is some angular coordinate of $S^5$ (it corresponds, in the notations
  of Appendix \ref{sec:ScalarSolutionsInAdS}, to $Y({\bf N})$ being a linear function of ${\bf N}$, {\it i.e.}
  $\Delta_S=1$). The simplest solution is $f = {1\over 5}(\log Z)Z\cos\theta$.
  It has nontrivial monodromy ${i\over 5}Z$.
  But we can add to it the expression
  ${1\over 5} \left((\log \overline{Z} ) Z + {1\over 2\overline{Z}}\sum X_i^2\right)\cos\theta$
  which is annihilated by $\square$ and cancels the monodromy.
  In general, exists solution with all logs being $\log(Z\overline{Z})$, no monodromy.
}\commentends

\vspace{5pt}
\noindent
Suppose that the monodromy were nontrivial. Let us consider the lowest order in $\epsilon$-expansion
where it be nontrivial. At the lowest order, it commutes with the undeformed action of $\bf g$.
Therefore, it must take values in a finite-dimensional representation (since the tensor product of
any number of $\cal H$ is still finite-dimensional). But finite-dimensional representations
are not unitary.

This argument implies, more generally, that the monodromy of finite-dimensional deformations is identity.

In Section \ref{sec:BoundarySMatrix} we will consider infinite-dimensional deformations,
with nontrivial monodromy.

\subsection{Lifting of $\Psi$ to superalgebra}\label{sec:ObstacleIsQuadratic}
Let $Q_{\rm ev}$ be the part of $Q$ (see Eq. (\ref{BRSTOperator})) involving only the ghosts
of {\em even} generators of ${\bf g}$ (essentially, all the ghosts of odd generators all put to zero).
The $\phi_{\rm bos}$ of Eq. (\ref{BosonicCocycle}) is annihilated by $Q_{\rm ev}$.
What happens if we act on $\phi_{\rm bos}$ with the full $Q$, including the terms containing odd indices?
Can we extend $\phi_{\rm bos}$ to a cocycle $\phi$ of ${\bf g}$?

To answer this question, let us look at the spectral sequence
corresponding to  ${\bf g}_{\rm ev}\subset {\bf g}$ \cite{FeiginFuchs}
It exists for any representation $V$. At the first page we have:
\begin{align}
  E_1^{0,1} \; = \;
  & H^1({\bf g}_{\rm ev}; V)
  \\
  E_1^{1,0} \;=\;
  & H^0({\bf g}_{\rm ev}; \mbox{Hom}({\bf g}_{\rm odd}, V)) \;=\;
    \mbox{Hom}_{{\bf g}_{\rm ev}}({\bf g}_{\rm odd}, V)
\end{align}
Our $\phi_{\rm bos}$ belongs to $E_1^{0,1}$. The first obstacle lives in
\begin{equation}
   E_1^{1,1}= H^1({\bf g}_{\rm ev}; \mbox{Hom}({\bf g}_{\rm odd}, V))
\end{equation}
We actually know that the SUGRA solution exist. Therefore this obstacle automatically vanishes.
But there is another obstacle, which arizes when we go to the second page. It lives in:
\begin{align}
  E_2^{2,0} \;=\;
  &
    H^2({\bf g},{\bf g}_{\rm ev};V)
    = H^2\left({\bf g},{\bf g}_{\rm ev};\mbox{Hom}({\cal H}^{\otimes 3} , \widehat{\cal H})\right)\;=
  \\
  \;=\;
  &
    H^2\left({\bf g},{\bf g}_{\rm ev};\mbox{Hom}({\cal H}^{\otimes 3} , {\cal H})\right)
\end{align}
We used the fact that relative
cochains are ${\bf g}_{\rm ev}$-invariant, therefore the cocycles automatically fall into the finite-dimensional
${\cal H}\subset\widehat{\cal H}$. In fact, this obstacle does not have to be zero, because
there is something that can cancel it. Remember that $\Psi$ is generally speaking not annihilated
by $Q$, but rather satisfies Eq. (\ref{MaurerCartan}). And, in fact, there is a nontrivial cocycle:
\begin{equation}\label{CocycleInS2HH}
   \psi \in H^1({\bf g}, \mbox{Hom}(S^2{\cal H}, {\cal H}))
\end{equation}
We conjecture
that the supersymmetric extension $\phi$ of $\phi_{\rm bos}$ indeed exists, but instead of
satisfying $Q\phi=0$ satisfies:
\begin{equation}\label{QPhiIsPsiPsi}
Q\phi = [\psi,\psi]
\end{equation}
\commentstarts{\small
  This conjecture should be verified by explicit computations, which we leave for future work.
  It may happen that the obstacle which would take values in the cohomology group of Eq. (\ref{CocycleInS2HH})
  actually vanishes for some reason. It seems that the computations done in  \cite{Aharony:2002hx}
  are not sufficient to settle this issue, because it was only done for one state (the beta-deformation
  constant in AdS)}\commentends

\vspace{10pt}
\noindent
We will now describe $\psi$ of Eq. (\ref{CocycleInS2HH}). 

\paragraph     {Step 1: construct an element of $H^1\left({\bf g}, \mbox{Hom}({\bf g}, {\cal H})\right)$}
Consider an element of $H^1\left({\bf g}, \mbox{Hom}({\bf g}, {\cal H})\right)$ corresponding
to the extension\footnote{Remember that $H^1\left({\bf g}, \mbox{Hom}(L_1,L_2)\right)$
is $\mbox{Ext}(L_1,L_2)$; it corresponds to the extensions \cite{Knapp} of $L_2$ by $L_1$.}: 
\begin{equation}
   0 \longrightarrow {\cal H}
   \longrightarrow {{\bf g}\wedge {\bf g}\over {\bf g}}
   \longrightarrow {\bf g} \longrightarrow 0
\end{equation}
(Remember that $\cal H$ is defined in Eq. (\ref{DefCalH}).)

\paragraph     {Step 2: compose it with an intertwiner $S^2{\cal H} \rightarrow {\bf g}$}
It was shown in \cite{Bedoya:2010qz} that there exists a $\bf g$-invariant map
\begin{equation}\label{FromS2HToG}
S^2{\cal H} \rightarrow {\bf g}
\end{equation}
--- we will review the construction of this map in Section \ref{sec:Intertwiner}.
Composing it with the element of $H^1\left({\bf g}, \mbox{Hom}({\bf g}, {\cal H})\right)$ we
get a class in $H^1({\bf g}, \mbox{Hom}(S^2{\cal H}, {\cal H}))$.

\subsection{Construction of the intertwiner $S^2{\cal H} \rightarrow {\bf g}$}\label{sec:Intertwiner}
We will now construct the intertwining operator in Eq. (\ref{FromS2HToG}).
\paragraph     {Algebraic preliminaries}
Suppose that we have an associative algebra $A$. For any $x_1\otimes\cdots\otimes x_{k} \in A^{\otimes n}$ consider their product:
\begin{equation} \mu(x_1\otimes\cdots\otimes x_{k}) = x_1 \cdots x_{k}\in A \end{equation}
In particular, take $A = \mbox{Mat}(m|n)$ — the algebra of supermatrices. Let us view the exterior product $\Lambda^k A = A\wedge\cdots\wedge A$ as a subspace in $A^{\otimes k}$.

\vspace{5pt}
\commentstarts{\small
  For any linear superspace $L$, there is a natural action of the symmetric group $S_n$ on the tensor
  product $L^{\otimes n}$. For example, when $n=2$, the transposition $\tau_{12}$ acts as:
  $\tau_{12} v\otimes w = (-)^{vw}w\otimes v$. The exterior product $\Lambda^n L$ is the subspace
  of $L^{\otimes n}$ where permutations act by multiplication by a sign of permutation. For example,
  for $n=2$ it is generated by expressions $v\wedge w = {1\over 2}(v\otimes w - (-)^{vw}w\otimes v)$.}
\commentends

\vspace{5pt}
\noindent
For any element $x_1\otimes\cdots\otimes x_{2k} \in \Lambda^{2k} A$ we define:
\begin{equation} \langle x_1\wedge\cdots\wedge x_{2k}\rangle = \mu(x_1\wedge\cdots\wedge x_{2k}) \end{equation}
We observe that:
\begin{align}
\mbox{STr}\langle x_1\wedge\cdots\wedge x_{2k}\rangle = 0
\\
\langle{\bf 1}\wedge x_2 \wedge \cdots \wedge x_{2k}\rangle = 0
\end{align}
Therefore, the operation $\langle\_\rangle$ defines a map:
\begin{equation} \Lambda^{2k}{\bf pl}(m|n) \rightarrow {\bf sl}(m|n) \end{equation}
We degine the ``split Casimir operator'':
\begin{equation}
   C = k^{ab} t_a\otimes t_b\;\in \; {\bf gl}(m|m)\otimes {\bf gl}(m|m)
\end{equation}
where $\{t_a\}$ are generators and $k^{ab}$ some coefficients, which we now define.
It satisfies:
\begin{equation}
   k^{ab}t_a\otimes [t_b,x] + (-)^{bx} k^{ab}[t_a,x]\otimes t_b = 0
\end{equation}
In particular, if we think of generators as matrices:
\begin{equation}
   [k^{ab} t_at_b\,,\,t_c] = 0
\end{equation}
In matrix notations:
\begin{equation}
   C^a{}_b{}^c{}_d = (-)^c\delta^a{}_d \delta^c{}_b
\end{equation}
Notice that for any matrix $X$:
\begin{equation}\label{in-the-middle}
     C^a{}_b{}^c{}_d X^b{}_c = \mbox{STr}(X) \; \delta^a{}_d
\end{equation}
We define:
\begin{equation} \delta x = k^{ab} t_a \otimes [t_b,x] = k^{ab} t_a \wedge [t_b,x] = (-)^{bx + 1} k^{ab}[t_a,x]\wedge t_b \end{equation}

\paragraph     {Description of $\cal H$}
In this language, the representation $\cal H$ in which beta-deformations transform consists of expressions:
\begin{align}
  \sum_i x_i\wedge y_i \in & {\bf sl}(4|4)\wedge {\bf sl}(4|4)
  \\
  \mbox{\tt\small modulo: }
                           & x\wedge {\bf 1}\simeq 0 
  \label{PEquivalence}\\  
  \mbox{\tt\small and } & \delta x\simeq 0
                          \label{DeltaEquivalence}
\end{align}

\paragraph     {Description of the intertwiner}
For $B_1$ and $B_2$ belonging to $\cal H$, we define:
\begin{align}
  f\;:\; & S^2{\cal H} \rightarrow {\bf g}
           \label{Intertwiner}
  \\   
  f(B_1\wedge B_2) \;=\;& \langle B_1\wedge B_2\rangle \; \mbox{mod} \; {\bf 1}
\end{align}
The correctness w.r.to the equivalences relation of Eq. (\ref{PEquivalence}) follows immediately.
It remains to verify the correctness w.r.to Eq. (\ref{DeltaEquivalence}). Indeed,
under the condition $[y,z]=0$ and $\mbox{STr}(y)=\mbox{STr}(z)=0$ we have:  
\begin{align}
  & 24 f(\delta x, y\wedge z)
    \;=\; 24 \left\langle \delta x \wedge y\wedge z \right\rangle
    \;=\; 24 \left\langle k^{ab}t_a\wedge [t_b,x] \wedge y\wedge z \right\rangle\;=\;
\nonumber\\ 
\;=\; & \phantom{+} (-)^{y(b+x) + bx}\langle k^{ab} t_a yx t_b z \rangle + (-)^{bx + by}\langle k^{ab} t_a xy t_b z \rangle +
\nonumber\\  
& + \; (-)^{yx + z(x+b) + bx}\langle k^{ab}yt_a zx t_b \rangle + (-)^{yx + zb + bx}\langle k^{ab}y t_a xz t_b \rangle\;-
\nonumber\\  
& - \; (-)^{zy} (z\leftrightarrow y)\;=
\nonumber\\  
\;=\; & 2\mbox{STr}(xy)\;z \;+\; (-)^{yz}2\mbox{STr}(xz)\;y \;-\;(-)^{zy} (z\leftrightarrow y)\;=\; 0
\end{align}
(We used Eq. (\ref{in-the-middle}).)
Therefore, we constructed a well-defined intertwining operator:
\begin{equation} S^2{\cal H} \longrightarrow {\bf g} \end{equation}
where ${\cal H} = {({\bf g}\wedge {\bf g})_0\over {\bf g}}$.

\paragraph     {Intertwiner maps $H^1\left({\bf g}, \mbox{Hom}({\bf g}, {\cal H})\right)$
  to $H^1\left({\bf g}, \mbox{Hom}(S^2{\cal H}, {\cal H})\right)$}
Our intertwining operator $f\;:\;S^2{\cal H}\rightarrow {\bf g}$ generates a short exact sequence:
\begin{align}
   & 0\rightarrow
     \mbox{Hom}({\bf g}, {\cal H}) \rightarrow
     \mbox{Hom}(S^2{\cal H}, {\cal H}) \rightarrow
     \mbox{Hom}(S^2_0{\cal H}, {\cal H}) \rightarrow 0
  \\
  &\mbox{\tt\small where } S^2_0{\cal H} =\mbox{ker} f
\end{align}
and therefore a long exact sequence:
\begin{align}
  0
  & \longrightarrow H^0\left({\bf g}, \mbox{Hom}({\bf g},{\cal H})\right)
    \longrightarrow H^0\left({\bf g}, \mbox{Hom}(S^2{\cal H},{\cal H})\right)
  \\  
  & \longrightarrow H^0\left({\bf g}, \mbox{Hom}(S_0^2{\cal H},{\cal H})\right)
    \longrightarrow
  \\
  & \longrightarrow H^1\left({\bf g}, \mbox{Hom}({\bf g},{\cal H})\right)
    \longrightarrow H^1\left({\bf g}, \mbox{Hom}(S^2{\cal H},{\cal H})\right)
    \longrightarrow \ldots
\end{align}
But $H^0\left({\bf g}, \mbox{Hom}(S_0^2{\cal H},{\cal H})\right) = 0$, therefore composition
with $f$ is an injective map 
$H^1\left({\bf g}, \mbox{Hom}({\bf g},{\cal H})\right)
\longrightarrow H^1\left({\bf g}, \mbox{Hom}(S^2{\cal H},{\cal H})\right)$.

We will now show that $H^0\left({\bf g}, \mbox{Hom}(S_0^2{\cal H},{\cal H})\right) = 0$, {\it i.e.}
there are no intertwining operators. Suppose that there {\em is} an intertwiner
\begin{equation}
\phi\;:\; S_0^2{\cal H} \rightarrow {\cal H}
\end{equation}
Let us compute it on a decomposable element $(x_1\wedge x_2)\bullet (y_1\wedge y_2) \in {\cal H}\bullet{\cal H}$.
Here $\bullet$ denotes the symmetrized tensor product.
The only way of contracting indices resulting in an antisymmetric tensor is:
\begin{equation}
\{x_2,y_1\}\otimes [x_1,y_2] - (-)^{(x_2+y_1)(x_1+y_2)}[x_2,y_1]\otimes \{x_1,y_2\}
\end{equation}
antisymmetrized over both $x_1\leftrightarrow x_2$ and $y_1\leftrightarrow y_2$.
(The terms like $[x,y]\otimes [x,y]$ belong to $S^2{\bf g}$ rather than $\Lambda^2{\bf g}$.)
But anticommutators are not allowed, because $\phi$ should be correctly defined
with respect to $x\simeq x + {\bf 1}$.

\subsection{Structure of $S^2{\cal H}$}
Let us denote:
\begin{equation}
   (S^2{\bf sl}(4|4))_{\rm STL}
\end{equation}
the subspace of $S^2({\bf sl}(4|4))$ consiting of elements $x\bullet y$ such that $\mbox{STr}(xy)=0$.
The map:
\begin{align}
  & S^2{\cal H}\rightarrow (S^2({\bf sl}(4|4)))_{\rm STL}
  \label{S2HtoS2sl}\\[5pt]   
  &(x_1\wedge x_2)\bullet (y_1\wedge y_2)\;\mapsto
  \nonumber\\  
   &\mapsto\;
   (-)^{x_2 y_1 + 1} [x_1,y_1]\bullet [x_2,y_2]
     + (-)^{x_1 y_1 + x_1 x_2} [x_2,y_1]\bullet [x_1,y_2]
     \nonumber
\end{align}
is an intertwiner. There is a map
\begin{align}
  \left[S^2{\bf sl}(4|4)\right]_{\rm STL} & \longrightarrow {\bf sl}(4|4)
  \label{S2sltosl}\\[5pt]
  x\bullet y & \mapsto xy
\end{align}
The composition of the map of Eq. (\ref{S2HtoS2sl}) and the map of Eq. (\ref{S2sltosl}), combined with the projector ${\bf sl}(4|4) \longrightarrow {\bf psl}(4|4)$,
equals to the map $f$ of Eq. (\ref{Intertwiner}):
\begin{align}
  & f\;:\;
    \left(\;
    S^2{\cal H}
    \longrightarrow
    \left[S^2{\bf sl}(4|4)\right]_{\rm STL}
    \longrightarrow
    {\bf sl}(4|4)
    \longrightarrow
    {\bf psl}(4|4)
    \;
    \right)
\end{align}
By definition $S^2_0{\cal H}=\mbox{ker}f$. This means that $S^2_0{\cal H}=\mbox{ker}f$ has some invariant
subspaces:
\begin{align}
  & \mbox{ker}\;
    \left(\;
    S^2{\cal H}
    \longrightarrow
    \left[S^2{\bf sl}(4|4)\right]_{\rm STL}
    \longrightarrow
    {\bf sl}(4|4)
    \;
    \right)
  \\   
  & \mbox{ker}\;
    \left(\;
    S^2{\cal H}
    \longrightarrow
    \left[S^2{\bf sl}(4|4)\right]_{\rm STL}
    \;
    \right)
\end{align}
This finer structure does not seem to be relevant for the leading term in the beta-function.

\subsection{Role of $\psi$ in anomaly cancellation}
Our construction of the cocycle as a product:
\begin{equation}
   H^1\left({\bf g}\;,\;\mbox{Hom}({\bf g}, {\cal H})\right)
   \otimes
   H^0\left({\bf g}\;,\;\mbox{Hom}({\cal H}\otimes {\cal H}, {\bf g})\right)
   \rightarrow 
   H^1\left({\bf g}\;,\;\mbox{Hom}({\cal H}\otimes {\cal H}, {\cal H})\right)
\end{equation}
suggests that it participates in anomaly cancellation.
It was explained in \cite{Bedoya:2010qz,Mikhailov:2012id} that at the level of the classical
sigma-model there is no reason for the parameter of the beta-deformation to have
zero internal commutator. From the point of view of the classical worldsheet, the beta-deformations
live in ${\bf g}\wedge {\bf g}\over {\bf g}$, and not necessarily in $({\bf g}\wedge {\bf g})_0\over {\bf g}$.
But at the quantum level, on the curved worldsheet,
the deformations with nonzero internal commutator suffer from one-loop anomaly.

This suggest the following anomaly cancellation scenario. Let us start with the linearized
{\em physical} ({\it i.e.} with zero internal commutator) beta-deformaion, and start
constructing, order by order in the deformation parameter $\epsilon$, the corresponding nonlinear solution.
The classical construction goes fine, but at the secondr order in $\epsilon$ we may encounter a one-loop anomaly
of precisely the right form to be cancelled by a non-physical beta-deformation. (``Nonphysical'' means
with non-zero internal commutator.) Then, we just add, with the coefficient $\epsilon^2$,
some nonphysical beta-deformation,  to cancel that anomaly.
But the subtlety is, that the extension of physical deformations by nonphysical:
\begin{equation}
   0
   \longrightarrow
   {({\bf g}\wedge {\bf g})_0\over {\bf g}}
   \longrightarrow
   {{\bf g}\wedge {\bf g}\over {\bf g}}
   \longrightarrow
   {\bf g}
   \longrightarrow
   0
\end{equation}
is not split. In other words, it is impossible to lift $\bf g$ back to ${{\bf g}\wedge {\bf g}\over {\bf g}}$
in a way preserving symmetries. In this sense, the anomally may break global symmetries.
In our language this means that the nontrivial $v_1$ of Eq. (\ref{ExpansionOfVectorField}) may be
induced by quantum corrections at the first order in $\alpha'$. 

But our conjecture is that the nontrivial $v_1$ is present already at the classical level
and its cohomology class participates in Eq. (\ref{CocycleInS2HH}). 

Both conjectures have to be settled by explicit computations, which we have not done.

\subsection{Outline of computation}\label{sec:OutlineOfComputation}

We believe that the best framework for actually computing $\Psi$ and proving the conjectured
Eq. (\ref{QPhiIsPsiPsi}) is the pure spinor formalism. This can be done using the
homological perturbation theory developed in \cite{Bedoya:2010qz,Benitez:2018xnh}. The basic idea is to consider
the deformation of the pure spinor BRST operator:
\begin{equation}
Q_{ps} = Q_{ps}^{(0)} + \epsilon Q_{ps}^{(1)} + \epsilon^2 Q_{ps}^{(2)} + \ldots
\end{equation}
The explicit expression for $Q_{ps}^{(1)}$ was obtained in \cite{Bedoya:2010qz,Benitez:2018xnh}. The next
step is to find $Q_{ps}^{(2)}$ such that $Q_{ps}$ is nilpotent up to terms of the order $\epsilon^{\geq 3}$.
This was done in \cite{Bedoya:2010qz,Benitez:2018xnh}, but only for a special class of deformations (essentially,
those leading to the integrable model, see \cite{Berenstein:2004ys,Bundzik:2005zg,Klimcik:2008eq,Delduc:2013qra,Hollowood:2014qma,Hoare:2015wia,Benitez:2018xnh}). The deviation from $\bf g$-covariance would arise for
non-integrable deformations, {\it i.e.} those cases where the $Q^{(2)}$ was {\em not} found in
\cite{Bedoya:2010qz,Benitez:2018xnh}.

\subsection{Other finite-dimensional deformations}
Besides beta-deformations, there are infinitely many other finite-dimensional
deformations \cite{Mikhailov:2011af,Mikhailov:2017uoh}.
The formalism developed in this paper should be also applicable to them.

\section{Comparison to boundary S-matrix}\label{sec:BoundarySMatrix}

\subsection{Periodic array of operator insertions}

Let ${\cal O}_1$ and ${\cal O}_2$ be two local operators, 
and $\rho_1$ and $\rho_2$ some c-number densities with support in sufficiently small
compact space-time regions.
Let us consider the following deformation of the action ({\it cf.} Eq. (\ref{SInv})) :
\begin{figure}[!htb]
     \center{\includegraphics[width=\textwidth]
     {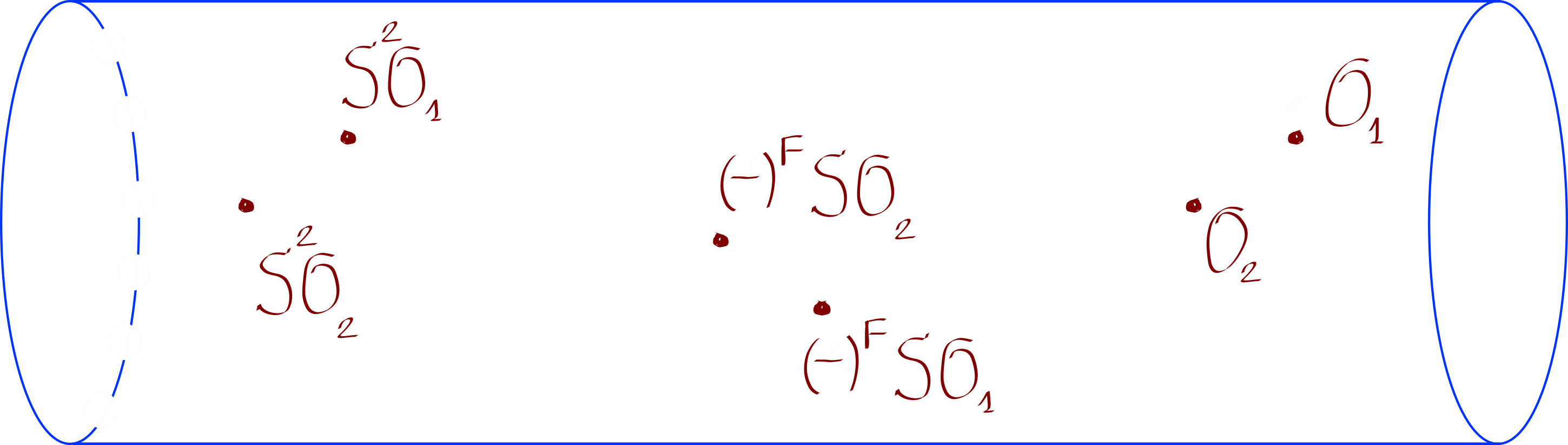}}
     \caption{\label{fig:TwoArrays} Two periodic arrays of insertions}
\end{figure}
\begin{equation}
   \delta S = \epsilon_1 \sum_{n=-\infty}^{\infty} \left((-1)^FS\right)^n \int d^4x \rho_1(x){\cal O}_1
   + \epsilon_2 \sum_{n=-\infty}^{\infty} \left((-1)^FS\right)^n \int d^4x \rho_2(x){\cal O}_2
\end{equation}
where $\epsilon_1$ and $\epsilon_2$ are two nilpotent coefficients.
This is, essentially, an infinite periodic array (designed to satisfy Eq. (\ref{SInv}))
of compactly supported deformations.

At the linearized level, {\it i.e.} assuming $\epsilon_1\epsilon_2=0$, the two terms
in the deformation transform in two infinite-dimensional representations, ${\cal H}_1$
and ${\cal H}_2$. But if we do not assume $\epsilon_1\epsilon_2=0$, then there will be
a term in the SUGRA solution proportional to $\epsilon_1\epsilon_2$, and it will {\em not}
transform in ${\cal H}_1\otimes {\cal H}_2$. We can consider the space of {\em all possible}
completions of linearized solutions to nonlinear solutions. The terms proportional to $\epsilon_1\epsilon_2$
form a linear space $X$, which, as a representation of $\bf g$, is an extension:
\begin{equation}
   0 \longrightarrow
   \left[\begin{array}{c}
         \mbox{\tt\small solutions of} \cr
         \mbox{\tt\small linearized equations}
      \end{array}\right]
   \longrightarrow
   X
   \longrightarrow
   {\cal H}_1\otimes {\cal H}_2
   \longrightarrow
   0
\end{equation}
Even if we restrict on ${\bf g}_{\rm even}$, there are such nontrivial extensions. The corresponding
cocycle:
\begin{equation}
   \psi\in H^1\left(
      {\bf g}_{\rm even}
      \;,\; \mbox{Hom}\left(
         {\cal H}_1\otimes {\cal H}_2\;,\;
         \left[\begin{array}{c}
                 \mbox{\tt\small solutions of} \cr
                 \mbox{\tt\small linearized } \cr
                 \mbox{equations}
               \end{array}
            \right]
      \right)
   \right)  
\end{equation}
is nontrivial, the average of $\psi(\partial/\partial t)$ over the period
(see Eq. (\ref{SecondDerivativeOfM})) is nonzero.

Let us insert, instead of an infinite array, just two operators: ${\cal O}_1$ and ${\cal O}_2$.
Consider the ``retarded'' solution excited by them. The waves will keep bouncing from the
boundary of AdS, interacting in the middle. Therefore the $\epsilon_1\epsilon_2$ part will
grow in global time like square of $t$. We consider this a complication. To simplify the analysis,
let us make four insertions (instead of just two):
\begin{equation}\label{FourInsertions}
   \epsilon_1{\cal O}_1 \;,\quad \epsilon_2{\cal O}_2 \;,\quad \epsilon_1(-1)^FS{\cal O}_1 \;,\quad
   \epsilon_2 (-1)^FS{\cal O}_2
\end{equation}
Then the terms linear in $\epsilon_1$, as well as the terms linear in $\epsilon_2$, will cancel
in the future. But, before they cancel, there will be some interaction, generating terms
proportional to $\epsilon_1\epsilon_2$. In the future, the $\epsilon_1\epsilon_2$-terms become
a solution of the free equation. This is the ``retarted'' solution generated by these insertions,
{\it i.e.} the one which is pure $AdS_5\times S^5$ in the past. 

\subsection{Monodromy {\it vs} boundary S-matrix}
Let us suppose that $\rho_1$ is a delta-function at the point $b_1$ on the boundary,
and $\rho_2$ delta-function at the point $b_2$. Genarally speaking, every point $b$ on
the boundary of AdS defines a Poincare patch, which can be defined as follows.
Consider the future of $b$, denote it ${\cal F}(b)$ (a subset of AdS). Notice that for any
$n>0$: $S^n {\cal F}(b) \subset {\cal F}(b)$. Consider the ``first fundamental domain'' of ${\cal F}(b)$
with respect to the action of $S$, {\it i.e.} the set of points $x\in {\cal F}(b)$ such $S^{-1}x\notin {\cal F}(b)$.
This is the Poincare patch ${\cal P}(b)$ corresponding to $b$ (the beige area on Figure \ref{fig:defS}). For the retarded solution
corresponding to the insertions (\ref{FourInsertions}) all the interaction happens
inside ${\cal P}(b_1)\cap {\cal P}(b_2)$.

This implies that the average of $\psi(\partial/\partial t)$, in the sense of Eq. (\ref{SecondDerivativeOfM})
can be computed as the integral 
over ${\cal P}(b_1)\cap {\cal P}(b_2)$.
In fact, since the boundary-to-bulk propagator has support on the light cone, see Eq. (\ref{BoundaryToBulkPropagator}), the integral is supported on $\partial{\cal P}(b_1)\cap \partial{\cal P}(b_2)$.
The integrand is the retarded propagator times triple-interaction vertex.

On the other hand, in the definition of the boundary S-matrix \cite{Witten:1998qj} the integration of the
interaction vertex is over the whole Euclidean AdS. It is not clear to us how these two definitions
are related.

\subsection{Normalizable and non-normalizable contributions to monodromy}\label{sec:NormalizableAndNonNormalizableMonodromy}
Notice that $\partial{\cal P}(b_1)\cap \partial{\cal P}(b_2)$ goes all the way to the boundary, therefore
there is no reason why the monodromy would be a normalizable solution.
However, the non-normalizable part is due to waves bouncing back and forth in AdS reflecting
from the boundary. Therefore all the non-normalizable terms can be damped by making adjustments,
of the order $\epsilon_1\epsilon_2$, of the boundary conditions. In other words, correcting
the defining Eq. (\ref{FourInsertions}) by adding some operators of the order $\epsilon_1\epsilon_2$.
Then, the monodromy of the modified array will be a normalizable solution. We used this in
Section \ref{sec:TrivialMonodromy}.

\section{Discussion and open questions}\label{sec:Discussion}

A general theme of AdS/CFT is comparison of field theory computations with supergravity
computations. The analysis of the present paper is incomplete, and potentially leaves
a {\em mismatch} between field theory and supergravity.
Indeed, on the field theory side we use the formalism
of Sections \ref{sec:IntroRenorm}, \ref{sec:ActionOfGOnDefs}, \ref{sec:GeometricaAbstraction}.
While on the supergravity side, we use the formalism of Sections \ref{sec:GeometricaAbstractionAdS},
\ref{sec:NormalFormA}, which is similar but different.

\subsection{Is it true that symmetries of QFT naturally act on the space of its deformations?}

It is essential for our reasoning, that there is a {\em natural action} of the symmetries
of QFT on the space of its deformations, Section \ref{sec:ActionOfGOnDefs}.
This action should be natural, {\it i.e.} should not depend on how we describe deformations.
Strictly speaking, our reasoning in Section \ref{sec:ActionOfGOnDefs} used a particular way
of thinking about the deformations. Therefore, we are in danger of using an unnatural definition.

\subsection{Gauge group is not an invariant}

Renormgroup invariants in QFT match certain cohomological invariants of the action of
the group of gauge transformations
of supergravity, as described in Section \ref{sec:NormalFormA}.
But gauge group is not actually an invariant of the theory
\footnote{Gauge transformations characterize the redundancy of the given Lagrangian desciption
of the theory. A different Lagrangian descriptions of the same theory can have slightly different
gauge symmetries.}. Therefore, we are in danger of being non-invariant.
However, the invariants which we describe in Section \ref{sec:NormalFormA}
actually depend, in some sense, on the cohomology of the gauge transformations.
Our construction uses certain ``flabbyness'' of the algebra of gauge transformations,
essentially allowing to reduce the cohomologies to those of $\bf g$ (using the Shapiro's lemma).
We hope that the cohomologies {\em are} invariant.

\subsection{Maybe there is no mismatch}

If our conjectures in Section \ref{sec:NormalFormA} hold, namely:
\begin{itemize}
\item Exists $M_{gauge-fixed}$ and
\item Solution of MC Equation does not depend on the choice of $M_{gauge-fixed}$
\end{itemize}
then the invariants of Section \ref{sec:NormalFormA} just match the invariants on the QFT side,
so there is no discrepancy.
We think that this is, most likely, what actually happens. 

\subsection{Computation of the MC invariants}

As we already mentioned in Section \ref{sec:OutlineOfComputation}, an important
open question is to develop the homological perturbation theory of \cite{Bedoya:2010qz,Benitez:2018xnh}
and to actually compute the Maurer-Cartan invariants which we defined.

\section*{Acknowledgments}
We want to thank Alexei Rosly for many useful discussions.
This work was supported
in part by FAPESP grant 2014/18634-9 ``Dualidade Gravitac$\!\!,\tilde{\rm a}$o/Teoria de Gauge''.
and
in part by RFBR grant RFBR 18-01-00460 ``String theory and integrable systems''.

\appendix

\section{AdS notations}\label{sec:AdSNotations}

\subsection{Embedding formalism}\label{sec:EmbeddingFormalism}

We here consider massless Laplace equation in $AdS_D$.
We realize $AdS_D$ as a hyperboloid in ${\bf R}^{2,D-1}$ parametrized by coordinates $Z,\overline{Z},X^1,\ldots,X^{D-1}$.
The equation of the hyperboloid is: 
\begin{equation}\label{Hyperboloid}
   |Z|^2 - \vec{X}^2 =  R^2
\end{equation}
\begin{align}
 &2{\partial\over\partial Z}{\partial\over\partial\overline{Z}}
        - {\partial\over\partial\vec{X}}{\partial\over\partial\vec{X}}
        \;=\;
        {1\over R^D}{\partial\over\partial R}R^D{\partial\over\partial R} + {1\over R^2}{\bf L}\;=
        \\\;=\;
 &{1\over R^2}
        \left(
              (\Delta + D - 1)\Delta + {\bf L}
   \right)
\end{align}
where  $\Delta = R{\partial\over\partial R}$ and ${\bf L}$  is the Laplace operator on $AdS_D$.
Therefore, on harmonic functions:
\begin{equation}
   {\bf L} = - \Delta(\Delta + D - 1)
\end{equation}
Our space-time is not just $AdS_D$, but $AdS_D\times S^D$. The formulas for Laplace operator on the sphere
are completely analogous. To distinguish between AdS and sphere, we use indices $\rm A$ and $\rm S$:
${\bf L}_{\rm A}$, $\Delta_{\rm A}$, ${\bf L}_{\rm S}$, $\Delta_{\rm S}$. The total Laplace operator
on $AdS_5\times S^5$ is:
\begin{equation}
   \square = {\bf L}_A - {\bf L}_S
\end{equation}
Therefore, for the scalar function to be harmonic in $AdS_D\times S^D$:
\begin{equation}
   \Delta_{\rm A}(\Delta_{\rm A} + D - 1) \;=\;\Delta_{\rm S}(\Delta_{\rm S} + D - 1) 
\end{equation}
This means:
\begin{equation}
\mbox{\tt\small either } \Delta_{\rm A} = \Delta_{\rm S} \quad \mbox{\tt\small or } \Delta_{\rm A} = - \Delta_{\rm S} - (D-1)
\end{equation}

\subsection{Solutions of wave equations}\label{sec:ScalarSolutionsInAdS}

Consider the following family of scalar field profiles, parameterized by a real parameter $\alpha$:

\begin{equation}
   \phi_{\alpha} = {f(\vec{X}) + \sum_{n=1}^{[\mbox{deg}f\,/2]} (Z\overline{Z})^n f_{n,\alpha}(\vec{X})\over Z^{{\rm deg}f + \alpha}}
   \end{equation}
where $f_{n,\alpha}$ can be determined recursively from:
\begin{align}f_{n,\alpha}(\vec{X}) \;=\;
 &- {n ({\rm deg}f_{(n-1),\alpha} + \alpha - n)\over 2} {\partial\over\partial\vec{X}}{\partial\over\partial\vec{X}}
               f_{(n-1),\alpha}(\vec{X})\\f_{0,\alpha}(\vec{X}) \;=\;
 &f(\vec{X})\end{align}
This solves the wave equation in ${\bf R}^{2,D-1}$
\begin{equation}
   \left(
         2{\partial\over\partial Z}{\partial\over\partial\overline{Z}} -
         {\partial\over\partial\vec{X}}{\partial\over\partial\vec{X}}
         \right)
   \phi_{\alpha} = 0
\end{equation}
Therefore:
\begin{equation}\label{CasimirOnPhiAlpha}
   {\bf L}_{\rm A} \phi_{\alpha} = \alpha(D - 1 - \alpha)\phi_{\alpha} 
\end{equation}

\paragraph     {Massless scalar in $AdS_D$}
To solve the wave equation on $AdS_D$, we take $\alpha=0$:
\begin{equation}
   {\bf L}_{\rm A} \phi_0 = 0
\end{equation}

\paragraph     {Massless scalar in $AdS_D\times S^D$}
Let us parametrize $S^D$ by a unit vector ${\bf N}\in {\bf R}^{D+1}$.
Suppose that the $S^D$ dependence is a harmonic polynomial $Y({\bf N})$
of order $\Delta_{\rm S}$. We must either take $\alpha = -\Delta_S$ or $\alpha = \Delta_S + D - 1$. The solution is:
\begin{equation}
\phi_{\alpha}(Z,\overline{Z},\vec{X}) Y({\bf N})
\end{equation}

\paragraph     {Inhomogeneous equations, appearence of log terms}
Consider the equation with nonzero right hand side:
${\bf L}_{\rm A}\phi = f$. When $f$ is proportional to $\phi$, the logarithmic terms appear. Indeed:
\begin{equation}
   {\bf L}_{\rm A} \left. {\partial\over\partial\alpha}\right|_{\alpha=0} \phi_{\alpha} = - (D-1)\phi_0
\end{equation}
and therefore:
\begin{equation}\label{LInvOnPhi0}
   {\bf L}^{-1} \phi_0 = - {1\over D-1}\left. {\partial\over\partial\alpha}\right|_{\alpha=0} \phi_{\alpha}
\end{equation}
This expression contains $\log Z$.
A somewhat special case is the equation:
\begin{equation}
   {\bf L}_{\rm A} \phi_1 = 1
\end{equation}
The solution is:
\begin{equation}\label{LogZBarZ}
   \phi_1 = {1\over 2(D-1)} \left.{\partial\over\partial\alpha}\right|_{\alpha = 0}
   \left(
      {1\over Z^{\alpha}} + {1\over \overline{Z}^{\alpha}}
   \right)
   = {1\over 2(D-1)}\log (Z\overline{Z})
\end{equation}
One can think of $\phi_{\alpha}$ as a family, parametrized by $\alpha$, of field profiles,
taking values in a different representation for each $\alpha$. (All these representations
are subspaces of one large space.) The value of the Casimir operator ${\bf L}_{\rm A}$ is given
by Eq. (\ref{CasimirOnPhiAlpha}). When $\alpha = 0$ it is zero.
From this point of view, Eq. (\ref{LInvOnPhi0}) is a particular case of the following
general construction. Suppose that we have an operator $\bf L$ acting on a representation space $V$ of $\bf g$,
commuting with $\bf g$,
and $V$ is a continuous direct sum  of subrepresentations $V_{\alpha}$ parametrized by a parameter $\alpha$,
such that the restriction of ${\bf L}$ on each $V_{\alpha}$ is the multiplication by $\alpha$.
For $v_0\in V_0$, we want to find $w$ such that $Lw = v_0$. Consider a 1-parameter family
of vectors $v(\alpha)\in V_{\alpha}$ such that $v(0)=v_0$. Then
$w = \left.{d\over d\alpha}\right|_{\alpha=0}v(\alpha)$. We only need a 1-jet of
the family. If it is possible to find a map:
\begin{equation}
   V_0 \rightarrow \mbox{\tt\small the space of 1-jets of paths in }V \mbox{ \tt\small passing through } V_0
\end{equation}
commuting with 
with the symmetry, then the equation $Lw = v_0$ can be solved in a covariant way:
\begin{equation}
w = \left.{d\over d\alpha}\right|_{\alpha=0}v(\alpha)
\end{equation}
For example, if $V$ were equipped with a metric, we could pick for each $v_0$ the path
going through $v_0$ with the velocity perpendicular to $V_0$. But in our context,
there is no invariant metric, and there is no ${\bf g}$-covariant invertion of ${\bf L}$.

We can construct a sequence of $t$-independent functions: 
\begin{align}
  \phi_0 \;=\;
  & 1
  \\   
  \phi_1 \;=\;
  & \log(Z\overline{Z})
  \\  
  {\bf L}_{\rm A} \phi_n \;=\;
  & \phi_{n-1}
    \label{BasicFunctions}
\end{align}
They all depend only on $Z\overline{Z}$ and grow near the boundary of AdS
as powers of $\log(Z\overline{Z})$.

\subsection{Functions participating in the perturbative expansion of nonlinear beta-deformation}
\label{sec:BasicFunctions}
We expect that nonlinear beta-deformation (and other finite-dimensional deformations)
is expressed in terms of
functions $\phi_n$ of Eq. (\ref{BasicFunctions}) and their derivatives w.r.to $Z$ and $\bar{Z}$,
multiplied by polynomials of $\vec{X}$ and rational functions of $Z,\overline{Z}$.


\def\cprime{$'$} \def\cprime{$'$}
\providecommand{\href}[2]{#2}\begingroup\raggedright\endgroup

\end{document}